\documentclass[10pt,a4paper]{article}
\usepackage[latin1]{inputenc}
\usepackage{amsmath}
\usepackage{amsfonts}
\usepackage{amssymb}
\usepackage{graphicx}
\usepackage{hyperref}
\title{Emergence in Solid State Physics and Biology}

\author{George F R Ellis\\
Mathematics Department, University of Cape Town}
\begin{document}
	\maketitle
\begin{abstract}
	There has been much controversy over weak and strong emergence in physics and biology.  As  pointed out by Phil Anderson in many papers,   the existence of broken symmetries is the key to emergence  of properties in much of solid state physics.  By carefully distinguishing between different types of symmetry breaking and  tracing the relation between broken symmetries at micro and macro scales, I demonstrate that the emergence of the properties of  semiconductors is a case of strong emergence. This is due to the existence of quasiparticles such as phonons. Furthermore time dependent potentials enable downward causation  as in the case of digital computers. Additionally I show that the processes of evolutionary emergence of living systems is also a case of strong emergence, as is the emergence of properties of life out of the underlying physics. A useful result emerges: standard physics theories  
	 and the emergent theories arising out of them 
	 are all effective theories that are equally valid. 
\end{abstract} 
\tableofcontents
\section{Emergence and Context}\label{sec:EFT_Intro}
A major concern in the interplay between science and philosophy is whether emergence is strong or weak when higher level properties  of a system emerge out of the properties of its constituent parts.

\subsection{Weak and Strong emergence}\label{sec:weak_strong}
David Chalmers \cite{Chalmers_2000}  defines weak and
 strong emergence as follows
\begin{itemize}
	\item \textbf{Weak Emergence of phenomena} We can say that a high-level phenomenon is weakly emergent with respect to a low-level
	domain when the high-level phenomenon arises from the low-level domain, but truths
	concerning that phenomenon are unexpected given the principles governing the low-level
	domain ...  It often happens that a high-level phenomenon is	unexpected given principles of a low-level domain, but is nevertheless deducible in principle
	from truths concerning that domain.
	\item \textbf{Strong Emergence of phenomena} We can say that a high-level phenomenon is strongly emergent with respect to a	low-level domain when the high-level phenomenon arises from the low-level domain, but
	truths concerning that phenomenon are not deducible even in principle from truths in the
	low-level domain. 
\end{itemize} 
I will accept those definitions; the latter is the concern of the body  of paper. Note that in dealing with emergence from a base theory, one will always of necessity make approximations of various kinds, or take limits of the underlying theory. The key issue raised by Chalmers' definition of strong emergence is the following: 
\begin{itemize}
	\item \textbf{Strong and Weak 
		 Derivation of Phenomena} Given that approximations or limiting procedures will be used in determining emergent phenomena, (i) are the variables used in this process deducible in principle from the lower level properties alone, together with a generic procedure for deducing higher level phenomena? Or (ii) are they  deducible from the lower level properties only when one adds  concepts or variables that are derived from our knowledge of the nature of the higher level emergent domain? The former case is strong derivation, indicating weak emergence,  and the latter case is weak derivation, indicating strong emergence. 
	 \end{itemize}
The point is that in case (ii), the lower level simply does not by itself imply the relevant higher level concepts or variables.
This is key: proposed interlevel bridge laws do not determine the higher level variables, but rather depend on them, as when \cite{Butterfield 2014} states one needs  empirical input from the higher level theory to determine such bridge laws. Strong upward emergence should determine these higher variables strictly from lower level concepts together with appropriate limits, approximations, and `upscaling' procedures (such as averaging)  alone. 

However `phenomena' is (deliberately) a very broad concept: it could mean electrical conductivity or rigidity, galaxies or planets, life or amoebae or giraffes, human beings or consciousness.   
This discussion can be extended to a related but more specific concept:
\begin{itemize}
	\item \textbf{Strong Emergence of Specific Outcomes} We can say that a high-level specific outcome is strongly emergent with respect to specific	low-level data when the high-level outcome arises from the low-level data, but the details of  that outcome is not deducible even in principle from data in the low-level domain alone. 
\end{itemize} 
Claims that this cannot occur are sometimes used in arguments against strong emergence; I will deal with this in Section \ref{sec:determine}.

\paragraph{Higher Level Variables}
As just indicated, a key point in the case of strong emergence is the need for ``new concepts'' at  higher levels \cite{Anderson_72}. Whenever such occur, this	is reason enough to think that the phenomena at the higher level are strongly emergent.  But the issue then is, Why are these new concepts
	indispensable? Why would it not in principle be possible to translate them with the help of lower level
	concepts? If there  is a ``dictionary'' relating different concepts at different levels, then emergence does
	not seem to introduce anything new. It rather seems to point to our limited ability to express
	complex emergent phenomena in terms of lower-level concepts. 
	
 I will show how such higher level variables do indeed occur as I consider strong emergence in various cases. In particular, this arises because of the issue of Multiple  Realisation discussed in Section \ref{sec_multiple}. I will consider the possibility of such a dictionary in Section \ref{Bio_variables} and  give examples of irreducible higher level variables in  Section \ref{sec:determine}.
  

\subsection{Contexts}\label{sec:contexts}
One may be interested in emergence in regard to broad areas of science such as physics, biology, or the mind/brain complex.  Indeed a huge part of the literature on emergence is concerned with the latter case. However there are very interesting issues as regards emergence in the case of physics as well.   

Given a choice of broad area, emergence may be rather different in different sub-fields in that area. 
If one is dealing with physics, then in the cases of particle physics or nuclear physics, use of Effective Field Theories (EFTs) \cite{Hartmann 2001} \cite{Castellani_02} \cite{LM} is a powerful method. This often shades over into renormalisation group methods \cite{Wilson 1982}, as for example in \cite{Burgess_2007}. 
However 
 I will deal with  somewhat different contexts of emergence: firstly solid state physics, giving details of how this works out for transistors in digital computers, and then broadly the case of biology.

\paragraph{Solid state physics}
In the case of solid state physics, an interesting situation arises as regards the coming into being  of the relevant crystals. Crystallization  is a case of weak emergence, arising due to energy minimisation as the crystal forms. However when either metals or semiconductors occur, they only exist because of the very specific manufacturing processes required to bring them into being; they are the outcome of purposeful design and manufacture \cite{Bronowski 2011} \cite{Mellis}. This is obvious from the fact that neither metals nor semiconductors occur naturally.

Once they are in existence, their properties may be strongly emergent for  two quite different reasons.
 In cases of everyday emergence such as  electrical and optical properties of metals and semiconductors, as well as in more exotic cases such as superconductivity, broken symmetries are the key, as forcefully pointed out by Phil Anderson \cite{Anderson_72} \cite{Anderson_1981_symmetry} \cite{Anderson_89} \cite{Anderson_94}. The emergent lattice structure in semiconductors leads to the existence of quasi-particles that inextricably link the lattice and electron levels via a  quantum physics effect of interlevel wave-particle duality. This  will be the focus of the first part of this paper. 
 I will not consider topological  effects \cite{Kvorning_2018}, such as occur in the Fractional Quantum Hall Effect, topological insulators, and some cases in Soft Matter Physics, even though they are excellent examples of strong emergence \cite{McLiesh_et_al_2019}. Neither will I deal with defect structures which lead to symmetry breaking with a  complex topological nature (\cite{Anderson 1984}:54-67). 


\paragraph{Biology} In the  case of biology, the coming into being over long periods of time of species is a case of strong emergence because of the contextual nature of the selection process \cite{Campbell_1974} \cite{Mayr 2001}. Then functioning at the molecular biology level is a case of strong emergence of properties due to the breaking of symmetry by molecular structure,  allowing time dependent constraints in the underlying Hamiltonian \cite{Ellis_Kopel_2019}.

\subsection{Downward causation and equality of levels}\label{sec:intro-down_equal}
There are two key points in the discussion. Firstly, I will claim  strong emergence is the result of a combination of upward and downward causal effects.  It is the downward influences that enable higher levels to shape lower level dynamic outcomes in accord with higher level emergent dynamics and needs. 
The existence of such downwards effects is denied by many physicists, but has been argued for strongly in the case of biology \cite{Campbell_1974}  \cite{Noble2008_Music} and in science more generally \cite{Ellis_2016}, including e.g. the case of structure formation in cosmology on the one hand and the functioning of digital computers on the other. In this
paper I give arguments for existence of downward causal effects in solid state physics (Section \ref{sec:downward_diachronic_synchronic}) as well as in biology (Section \ref{sec:biology_diachronic}). 

As a consequence of this confluence of effects, Effective Theories (ETs)  (which may or may not be Effective Field  Theories) emerge at each level of the hierarchy, characterised by a vocabulary and dynamics applicable at that level. The issue then is, Is there a fundamental level that is more basic and has more causal power than all of the others?

It seems to be assumed by most physicists that this is indeed the case - there is an underlying most basic level (a `Theory of Everything', or TOE) from which all else follows. 
 However the problem is that we do not have such a well-defined and tested fundamental theory to which all else could in principle be reduced. We do not even have a proof that such a level exists. That supposition is an unproven philosophical assumption, which may or may not be true. In any case alleged reduction to an ill-defined theory is problematic.    
 
 As a consequence, all the theories that physicists deal with in practice are Effective Theories \cite{Castellani_02}, assumed to emerge from that unknown underlying TOE. None of them is \textit{the} underlying theory. And in consequence physicists use whatever theory is convenient for a particular application, not worrying about the fact that it is not the alleged TOE. Various levels may be chosen as the `lowest' level in any particular study. 
 
 Denis Noble formalized this feature in the case of biology by his `Principle of Biological Relativity'  \cite{Noble 2012}, stating that all emergent levels in biology should be regarded as equally causally efficient: none of them has supremacy over the others. In this paper (Section \ref{sec:equal_valid}) I extend that statement to a generally applicable  \textit{Principle of Equal Causal Validity}: all emergent levels, whether physical or biological, are of equal causal validity in that they support an Effective Theory that may be considered to be valid at that level, and is not trumped by effective  laws at any other level. And this is the case because of the downward causation that occurs equally with upward causation. 

\subsection{This paper}\label{sec:this_paper}
In the following, Section \ref{sec:conceptual}  introduces some key conceptual distinctions that underlie the analysis in the rest of the paper. Section \ref{sec:Emerge_Physics} considers issues relating to emergence in solid state physics: the significance of broken symmetries, problems with purely bottom up derivations of properties, and the nature of quasiparticles such as phonons. 

Section \ref{sec:Strong_emerge} 
demonstrate that strong emergence occurs in condensed matter physics (\S \ref{sec:SSB(M)}). 
 Section  \ref{sec:downward_diachronic_synchronic} shows that standard microdynamics \textbf{m} is causally incomplete unless one includes downwards effects from higher levels. Section \ref{sec:details} shows how this all works in the case of transistors in digital computers. Section \ref{sec_multiple} comments on the multiple realisability of higher level states by lower level states, which is a key aspect of strong emergence. 
	

Section \ref{sec:emerge_life} considers emergence and life.  Life involves function and purpose that are emergent characteristics that cannot be expressed in terms of lower level variables (\S \ref{sec:purpose}). The existence of life is a case of strong emergence due to natural selection (\S \ref{sec:natural_select_emerge}). Furthermore properties of all living systems are strongly emergent because of the symmetry breaking due to the existence of molecular structure (\S \ref{sec:biology_diachronic}), allowing downward causation via time dependent constraints.  Thus life is strongly emergent for multiple reasons  (\S \ref{sec:Strong_Emerge_Biol}). 

Section \ref{sec:conclude}  discusses essentially higher level variables (\S \ref{sec:higher_var}) the important result that all emergent levels are equally causally valid in terms of supporting an effective theory (\S \ref{sec:equal_valid}). 
    
\section{Conceptual Issues}\label{sec:conceptual}
 This paper uses the idea of an Effective Theory (Section \ref{sec:EffectiveTheories}). The conceptual bases of the present paper are twofold: first, characterizing  different aspects of emergence (\ref{sec:emerge_types}); second, characterizing different kinds of symmetry breaking (Section \ref{sec:kinds_of_symmetry_breaking}). 
 
 \subsection{Effective Theories}\label{sec:EffectiveTheories}
 Elena Castellani  gives this definition \cite{Castellani_02}: 
 \begin{quote}
 	``\textit{An effective theory (ET) is a theory which `effectively' captures what is physically relevant in a given domain, where `theory' is  a set of fundamental equations (or simply some Lagrangian) for describing some entities, their behaviour and interactions... More precisely,  an ET is an appropriate description of the important (relevant) physics in a given region of the parameter space of the physical world. ... It is therefore an intrinsically approximate and context-dependent description.}''
 \end{quote} 
 A key result is that all emergent  levels are equally causally effective  in the sense that the dynamics of every level \textbf{L} we can deal with in an empirical way is described by  an Effective Theory $\textbf{EF}_\textbf{L}$ at that level (see (\ref{eq:effective_laws})) that is as valid as the effective theories at all other levels (\S\ref{sec:equal_valid}). There is no privileged level of causality \cite{Noble 2012}. 
 
 \subsection{Different aspects of emergence}\label{sec:emerge_types}
 I distinguish two different aspects of the nature of emergence as follows.
\begin{itemize}
	\item Emergence \textbf{E} of system  from its components. For example, the emergence \textbf{E} of nuclei out of protons and neutrons, of water or a  metal or  hemoglobin molecules out of the underlying nuclei and electrons, or of a human body out of its constituent cells. The issue of  \href{https://en.wikipedia.org/wiki/Phase_transition}{phase transitions} is important here. These occur when a major change in the emergent state takes place, such as the transition of  water from a liquid to a gaseous state when boiling occurs.
	\item Emergence \textbf{P} of properties of the emergent system out of its underlying constituents once it has come into existence. How do properties of a nucleus arise out of the nature of its constituent neutrons and protons, and theirs out of the constituent quarks? How do rigidity or electrical conductivity or optical properties of a crystal, or chemical properties of a molecule, arise out of the underlying electrons, protons and neutrons? How do properties of  a cell in a human body arise out of properties of its underlying biomolecules? How does behaviour arise out of those cells?
	 \end{itemize}
The issue of whether emergence is strong or weak arises in both cases.
This paper is concerned with \textbf{P} rather than \textbf{E}, except in the case of biology where it matters what kinds of entities are present in the world today: \textbf{E} underlies \textbf{P} (Section \ref{sec:emerge_life}).  
\subsection{Different kinds of symmetry breaking}\label{sec:kinds_of_symmetry_breaking}
 Phil Anderson in his famous paper ``More is Different'' \cite{Anderson_72} and later writings \cite{Anderson 1984} \cite{Anderson_89} \cite{Anderson_94}  emphasized the crucial importance of symmetry breaking for emergence in condensed matter physics.  For the purposes of this paper, it is useful to distinguish three different types of symmetry breaking occurring in the context of  the relation of macro dynamics  \textbf{M} to microdynamics \textbf{m}.
\begin{itemize}
	\item \textbf{SSB}(\textbf{m}), Spontaneous Symmetry Breaking \textbf{SSB} occurring at the micro level $\textbf{m}$;   
	\item \textbf{SSB}(\textbf{M}), Spontaneous Symmetry Breaking \textbf{SSB} occurring at the macro level \textbf{M} due to the process  
	 \textbf{E} creating that level  from the micro level \textbf{m}, that is \textbf{E}: $\textbf{m} \rightarrow \textbf{M}$;
	\item \textbf{SB}(\textbf{NS}), Symmetry breaking \textbf{SB} in biology that has occurred due to Darwinian processes of natural selection \textbf{NS} leading to existence of specific kinds of molecules at the micro level \textbf{m} \cite{Wagner 2014}. 
\end{itemize}
These differences will be key below.

\section{
	Emergence and Solid State Physics}\label{sec:Emerge_Physics}
Solid state physics is a particular well-trodden area in which to investigate emergence of properties. This section sets the stage for that investigation in the next section.  

It considers in turn, the key feature of Broken Symmetries in solid state physics (Section  \ref{sec:Broken_symmetries}),  problems arising in trying to determine properties of solid state systems in a purely bottom up way (Section \ref{sec:bottom_up_problems}), and the key role of phonons in their properties (Section \ref{sec:Phonons}).   Section \ref{sec:downward} views this in another way: in terms of  Downward Emergence.

\subsection{Broken Symmetries}\label{sec:Broken_symmetries} 

Broken symmetries are key to emergence in condensed matter physics \cite{Anderson_72}, \cite{Anderson_1981_symmetry}.
\href{https://en.wikipedia.org/wiki/Spontaneous_symmetry_breaking}{
Spontaneous symmetry breaking} naturally leads to a class of emergent entities with broken symmetries.

 The simplest example is a crystalline state (\cite{Anderson_89}:587):
\begin{quote}
	\textit{``Why do we call the beautifully symmetric crystalline state a `broken symmetry? Because, symmetrical as it is, the crystal has less symmetry than the atoms of the fluid from which it crystallized: these are in the ideal case featureless balls, while the crystal has no continuous symmetry or translation symmetry}''.
\end{quote}
Thus the symmetry of a crystal breaks the translational invariance of the theory that underlies it,  and leads to the Bloch wave functions that are key to much of solid state physic. By Bloch's theorem, the electron wave functions in a crystal has a basis  of Bloch wave energy eigenstate. This feature underlies the nature of electronic band structures, and thus for example determines electric resistivity and optical absorption,  which are measurable phenomena.  
What this means is that physical properties  of the emergent state (the specific crystal structure) reach down to influence the motions of electrons moving in the lattice. That's a downward  effect underlying key physical properties of semiconductors \cite{Ellis_2016}. I return to this in \S \ref{sec:Phonons}.
\subsection{Problems with bottom up derivations}\label{sec:bottom_up_problems}
 \href{https://www.nobelprize.org/prizes/physics/2003/leggett/biographical/}{Anthony Leggett} in his article ``On the nature of research in condensed-state physics'' (\cite{Leggett_1992}, quoted in \cite{Drossell_2020}) writes as follows:\footnote{Basically the same is true in quantum chemistry, see \cite{Hunger 2006}.}
\begin{quote}
	\textit{``No significant advance in the theory of matter in bulk has ever come about through derivation from microscopic principles. (...) I would  confidently argue further that it is in principle (my emphasis) and forever impossible to carry out such a derivation. (...) The so-called derivations of the results of solid state physics from microscopic principles alone are almost all bogus, if `derivation' is meant to have anything like its  usual sense.}''
\end{quote}
Thus he is proclaiming strong emergence. He gives  Ohm's Law as a specific example:
\begin{quote}
	\textit{``Consider as elementary a principle as Ohm's law. As far as I know, no-one has ever come even remotely within reach of deriving Ohm's law from microscopic principles without a whole host of auxiliary assumptions ('physical approximations'), which one almost certainly would not have thought of making unless one knew in advance the result one wanted to get, (and some of which may be regarded as essentially begging the question)}'' \cite{Leggett_1992}.
\end{quote} 
Thus one often needs some additional and
	distinct assumptions in order to derive an effective theory. These assumptions are usually called
	``auxiliary assumptions'' and do not necessarily involve approximations. This is what
	happens when try we derive an effective theory in the bottom-up fashion,  
	 but need
	some ``extra information'' about the low energy degrees of freedom (e.g., some new symmetry
	principle) in order to do so. 
	
	Can one derive the results in a purely bottom up way? No. One can derive the micro-macro relation in an upward way provided one inserts some of the macro picture either into the effective Lagrangian  as a symmetry breaking term (see \S \ref{sec:Strong_emerge}), or into the micro-macro relation, as Weinberg does in his effective field derivation of superconductivity \cite{Weinberg 2000}: 
\textit{``instead of counting powers of small momenta, one must count powers of the departures of momenta from the Fermi surface. Without that ingredient, the derivation fails''}.   
The point then is that you can only derive the Fermi surface position from the macro theory. Thus it is not a purely bottom up derivation. If you try one, you will fail, as stated strongly by \cite{Laughlin_1999}, so it is weak derivation.\footnote{Reminder: I am using David Chalmer's definitions, see  Section \ref{sec:weak_strong}.} 


\subsection{Phonons and Quasiparticles}\label{sec:Phonons}
The problem in the bottom up derivation of emergent properties in solid state physics  is not just a question of not having enough computing power. It is a question of having the right concepts at hand. And you can't get those by studying the low level dynamics \textit{per se}, in most cases, because they involve high level concepts. 
 
The key feature is that  \href{https://en.wikipedia.org/wiki/Spontaneous_symmetry_breaking}{Spontaneous Symmetry Breaking} (\textbf{SSB}) takes place, so the symmetries of the equations are not shared by the solution \cite{Anderson 1984}. 
This leads to the  properties \textbf{P} of emergent highly ordered structures such as crystals that, through their ordered nature, lead to everyday properties such as stiffness and optical and electrical properties, but also can lead  
lead to complex behaviours such as  superconductivity and superfluidity. This emergence is an essentially quantum phenomenon with two key aspects. 

\paragraph{Interlevel Wave-particle duality}
The existence of \href{https://en.wikipedia.org/wiki/Quasiparticle}{quasiparticles} such as  \href{https://en.wikipedia.org/wiki/Phonon}{phonons} due to the broken symmetries of the emergent lattice structure is a situation  where they come into being at the lower level because they are dynamically equivalent to collective vibrations of a higher level  structure (the crystal lattice) (\cite{Ziman1979}:60, \cite{Goodstein 1985}: 154-155, \cite{Phillips2012}:172-175, \cite{Lancaster}). While they are vibrational modes of the lattice as a whole, and hence emergent entities, they nevertheless have particle like properties. Steven Simon expresses it like this (\cite{Simon 2013}:82-83):
\begin{quote}
	``\textit{As is the case with the photon, we may think of the phonon as actually being a particle, or we can think of the phonon as being a quantized wave. If we think about the phonon as being a particle (as with the photon) then we see that we can put many phonons in the same state (i.e., the quantum number n can be increased to any value), thus we conclude that phonons, like photons, are bosons}.''	
\end{quote}
Thus the lattice vibrations at the macro scale are dynamically equivalent to a particle at the micro scale. Adding them to the micro scale description as quasi-particles such as phonons, one now has a symmetry breaking micro theory that can be the basis of for 
derivation of a symmetry breaking macro theory. The phonons are an  essential element  of the micro theory (see \cite{Simon 2013}, pp. 82 on).  
This is an essentially quantum phenomenon: a form of the standard wave-partical duality of quantum physics.
\begin{quote}
	 \textbf{Wave-particle duality: There exists a \textbf{\textit{wave(macro)-particle(micro) duality}} in crystal structures which provides the crucial interlevel link in emergence of properties. 
}\end{quote}
This is key to what happens physically.
\cite{Guay and Sartenaer_quasiparticles} discuss in depth the philosophical implications of the existence of quasi-particles.

\subsection{Downward emergence}\label{sec:downward}
One can view this in another way. A crystal structure causes existence of phonons at the micro level, a case of downward emergence \cite{Franklin and Knox 2018}.\footnote{Called a ``Foundational
	Determinative Relation'' (FDR) by Carl Gillett, see  \cite{Gillett}.}  These lead to 
 the emergent properties that occur at the macro level in semiconductors  \cite{Ziman1979}, 
\cite{Grundmann_2010-Semiconductors}, 
\cite{Simon 2013}. Thus
downward effects shape the lower level (electron) dynamics due to the higher level (crystal) structure, these effects 
not being directly implied by the lower level interactions by themselves. 

\begin{quote}
\textbf{Downward emergence: The higher level context alters the lower level dynamics by introducing into it quasi-particles such as phonons that play a crucial role in solid state physics}
\end{quote}  
   But a lingering doubt remains: are they real, or fictitious?
Stephen Blundell gives the answer (\cite{Blundell 2019}:244),
  \begin{quote}
  	\textit{``So now we come to the key question: Are these emergent particles real? From the perspective
  	of quantum field theory, the answer is a resounding yes. Each of these particles emerges from a
  	wave-like description in a manner that is entirely analogous to that of photons. These emergent
  	particles behave like particles: you can scatter other particles off them. Electrons will scatter off
  	phonons, an interaction that is involved in superconductivity. Neutrons can be used to study the
  	dispersion relation of both phonons and magnons using inelastic scattering techniques. Yes, they
  	are the result of a collective excitation of an underlying substrate. But so are `ordinary' electrons  	and photons, which are excitations of quantum field modes.'}'
  \end{quote}
 The \href{https://en.wikipedia.org/wiki/Cooper_pair}{Cooper pairs} responsible for superconductivity are similarly downwardly emergent lower level effective variables, which can be regarded as produced either by electron-phonon interactions at the lower level, or by crystal distortions at the higher level. They would not exist were it not for specific kinds of crystal structure.  

\section{Strong Emergence in Solid State  Physics}\label{sec:Strong_emerge}
 This section is the heart of the argument on emergence in relation to solid state physics.  Section \ref{sec:foundations} clarifies  what will be assumed to be the underlying microphysics.  Section \ref{sec:SSB(m)} shows that while Spontaneous Symmetry Breaking at the micro scale (\textbf{SSB}(\textbf{m})) can lead in principle to symmetry breaking at the macro scale, this is not a significant feature of the emergence of properties in solid state physics in practice. 
 
 Section \ref{sec:SSB(M)} demonstrates that strong emergence  takes  place in solid state systems whose emergent properties are a result of Spontaneous Symmetry Breaking \textbf{SSB}(\textbf{M}) occurring  via emergence processes $\textbf{E}:\textbf{m} \rightarrow \textbf{M}$.  
 Although the macro dynamics \textbf{M} do not follow from the micro dynamics  \textbf{m} because they do not have the correct symmetries to allow this,  
one can nevertheless obtain an \textit{Effective Theory} \textbf{m'} at the microscale by introducing quasiparticles such as phonons. This implies downward causation takes place, because their existence depends on the existence of the crystal structure. This leads to the statement that all the well established  Laws of Physics are Effective Theories.  
 
 Section \ref{sec:downward_diachronic_synchronic} examines this feature of downward causation and its relation to the alleged Causal Completeness  of physics at the micro scale, and turns the usual argument on its head: I claim that microphysics \textbf{m} cannot in fact be causally complete, because it is unable by itself to lead by any coarse graining process to the correct macro dynamics \textbf{M}.
 Section \ref{sec:P(d)_solid_state} on  downward causation and effective laws 
  confirms this result. To give flesh to this story, Section \ref{sec:details} discusses the case of transistors in a digital computer.
    
 
 An underlying issue of importance is that multiple realisability of higher level structures and functions in terms of lower level structures and functions underlies downwards causation, and hence strong emergence. I  discussion this in Section  \ref{sec_multiple}.
 
 \subsection{Setting the Foundations}\label{sec:foundations}
 The issue is whether strong emergence takes place in physics. 
 \paragraph{Microphysics} To be clear about the context, what kind of micro theory \textbf{m} do people usually have in mind? Robert Bishop  clarifies as follows, in relation to patching physics and chemistry together \cite{Bishop_2005_patch}:
\begin{quote}\textit{
	``In quantum chemistry, one first specifies the fundamental physical
	interactions (electromagnetic, strong- and weak-nuclear, etc.), then enumerates the relevant
	particles and their properties (nucleon, electron, charge, mass, etc.). Next, one lists the pairwise
	interactions among the particles. Finally, one writes down the kinetic and potential energy operators
	and adds them to get the system Hamiltonian (an expression for the total energy of the system).
	With the Hamiltonian in hand, one then proceeds to derive the properties and behaviors of the
	chemical system in question.''}
\end{quote}
But what is the specific set of interactions one should take into account in \textbf{m}? For ordinary life, depending on the scale considered,  it will be Newton's Laws of Motion, together with the appropriate force laws. These are given by  Maxwell's equation of motion, together with Maxwell's equations for the electromagnetic field. One may need the Schr\"{o}dinger equation, or maybe Quantum Field Theory (QFT) and the Dirac equation.  

An interesting issue is gravitation. For ordinary laboratory scale physics we do not need Newton's Gravitational Laws, but if the scale is large enough, we may need to include Galilean gravity represented by an effective acceleration \textbf{g}. This is a symmetry breaking contextual term resulting from downward causation due to the existence of the Earth with a specific mass $M_E$ and radius $R_E$. This can have a significant effect on atomic scale phenomena, as in the case of \href{https://en.wikipedia.org/wiki/Atomic_fountain}{atomic fountains} underlying the existence of \href{https://en.wikipedia.org/wiki/NIST-F1}{cesium fountain} clocks. It will not play a role in the solid state physics considered here such as physics of transistors, although of course it does play an important role in biology and engineering. 

The above is what I will have in mind in referring to microphysics $\textbf{m}$ below.\footnote{This is made explicit by \cite{Laughlin_Pines_2000} in their equations [1] and [2]; they say ``Eqs. [1] and [2] are, for all practical purposes, the Theory of Everything for our everyday world.''}
 
\paragraph{Broken symmetries} As mentioned above, Anderson claims that the key to 
emergence in solid state physics is broken symmetries, 
which occurs via  Spontaneous Symmetry Breaking (\textbf{SSB}).
However there is a crucial issue here. \textbf{SSB} can take place because of the way the macrostructure \textbf{M} comes into existence. As discussed in Section \ref{sec:conceptual}, I call these cases \textbf{SSB}(\textbf{M}). 
But \textbf{SSB} can also take place at the micro level \textbf{m}.  I call this \textbf{SSB}(\textbf{m)}.  This distinction plays a key role in the discussion below.

\subsection{The case of \textbf{SSB}(\textbf{m})}\label{sec:SSB(m)}
Here 
the needed spontaneous symmetry breaking takes place at 
the micro scale, for example via the Higgs mechanism 
resulting from the Mexican Hat shape of the Higgs 
potential. Then macro symmetry breaking is in principle 
deducible in a bottom up way through coarse graining 
this symmetry-broken micro state. Weak emergence  
takes place. In notional terms, if the coarse 
graining operation \textbf{C} commutes with the symmetry 
\textbf{S}, then\footnote{Complexities can arise regarding 
the commutation here because of the ``$\neq$'' relation. 
The implication ``$\Rightarrow$'' here and in (\ref{eq:symm_coares}) should be read ``can imply'' rather than ``implies''.  This complication does not apply to (\ref{eq:not_possible}) because that implication does not use commutation of an inequality.}
\begin{equation}\label{key}
\{\textbf{SSB}(\textbf{m}): \textbf{S}(\textbf{m})\neq \textbf{m},\,\,   \textbf{C}(\textbf{m}) = \textbf{M},\,\, \textbf{CS}=\textbf{SC}\}\Rightarrow   \textbf{S}(\textbf{M}) \neq \textbf{M}.
\end{equation}
 However the known symmetry breaking mechanisms   producing their symmetry breaking effects at the microscale  level have limited higher level effects. While their knock-on effects reach up to scales relevant to the Standard Model of Particle Physics, they do not, in a daily life context,\footnote{I am excluding discussion of the context of the very early universe where cosmic inflation took place. In that case, \textbf{SSB}(\textbf{m}) had a major effect at macro scales.} reach up to the scale of atomic physics or higher, and so do not affect phenomena such as solid state physics or  chemistry or microbiology. 
The point is that the masses of
quarks and gluons  in the Standard Model (and hence the Higgs mechanism) are so much smaller than
 the mass of composite entities at larger distance scales (neutrons, protons, atoms,
etc.) because the Electroweak scale is 246 GeV 
while the  
neutron and proton mass is 940 MeV, a difference of 3 orders of magnitude. Therefore ``\textit{energies of relevance to nuclei can never
	reveal their quark-gluon substructure, it is simply irrelevant at these energies''} \cite{LM}. Consequently we conclude 
\begin{quote}
\textbf{Conclusion} \textbf{\textit{SSB(m) cases can produce the required symmetry breaking at higher levels M through weak  emergence. However while of fundamental importance in the standard model of particle physics, this is not relevant to situations such as the emergence of properties of  semiconductors and metals.} }	
\end{quote}
\subsection{The case of \textbf{SSB}(\textbf{M})} \label{sec:SSB(M)}
Here the needed symmetry breaking does not take place at the micro scale, but rather through emergence of the macro scale, due to interactions that minimise energy of the emergent structure; think of atoms crystallizing to form a crystal,  for instance. This is spontaneous emergence of the breaking of symmetry  \textbf{S} at the macro scale \textbf{M}, hence is a case of  \textbf{SSB}(\textbf{M}). In notional terms, if $\textbf{E}$ is the process of emergence, 
\begin{equation}\label{key}
\{\textbf{S}(\textbf{m})=\textbf{m}, \,\,\,\textbf{E}(\textbf{m}) = \textbf{M}\} \Rightarrow \textbf{M}: \,\,\textbf{S}(\textbf{M}) \neq \textbf{M}.
\end{equation}
The emergence process $\textbf{E}$ alters the macro symmetry by processes of energy minimisation based in the underlying microphysics $\textbf{m}$, and this is weak  emergence. But that is not the concern here. The issue is \textbf{P}: how do we determine the physical properties  of the emergent structure \textbf{M}, such as electrical conductivity,  once it exists? \\

As discussed above (\S \ref{sec:Phonons}),  existence of the macrostructure changes the context of the microphysics, resulting in effective interactions with quasiparticles such as phonons.\footnote{See the Blundell quote in Section \ref{sec:downward}.} Thus the emergent broken symmetry at the macro scale reaches down to affect conditions at the micro scale, thereby causing effective symmetry breaking at that scale. This produces from \textbf{m} an effective micro theory $\textbf{m'}$ that breaks the symmetry \textbf{S}, for example by including \href{https://en.wikipedia.org/wiki/Phonon}{phonons} or other \href{https://en.wikipedia.org/wiki/Quasiparticle}{quasi-particles} in the dynamics, and that therefore can produce the required symmetry-broken macro theory by coarse graining.  
In notional terms
	\begin{equation}\label{key}
	\textbf{SSB}(\textbf{M}) \Rightarrow \textbf{m} \rightarrow  \textbf{m'}:\,\,\textbf{S}(\textbf{m'}) \neq \textbf{m'}.
	\end{equation}
	 Typically this is done by introducing at the micro level a variable $\textbf{a}(\textbf{M})$ that is determined by  macro level conditions, and breaks the symmetry \textbf{S}:
	\begin{equation}\label{eq:downward}
	\{\textbf{m} \rightarrow \textbf{m'} = \textbf{m'}(\textbf{a}), \,\,\partial \textbf{m'}/\partial \textbf{a} \neq 0, \,\, \textbf{S}(\textbf{a}) \neq \textbf{a}\} \Rightarrow \textbf{S}(\textbf{m'}) \neq \textbf{m'}.
	\end{equation}
	This is the way that ETs for physics involving symmetry breaking can be derived: you apply a coarse graining process $\textbf{C}$ to $\textbf{m'}$, not $\textbf{m}$, and thereby obtain the macro theory \textbf{M} that breaks the relevant symmetry and accords with experiment: that is, 
	\begin{equation}\label{eq:symm_coares}
	 \{\textbf{C}(\textbf{m'})=\textbf{M}\} \,\,\Rightarrow\,\,\, \textbf{S}(\textbf{M}) \neq \textbf{M}.
	\end{equation}  
	How this works out in the case of functioning of transistors is explained in Section 4 of \cite{Ellis_Drossel_computers}, see Section \ref{sec:details} below. In some approaches to quantum chemistry, this is done via the \href{https://en.wikipedia.org/wiki/Born%E%80%93Oppenheimer_approximation}{Born-Oppenheimer} approximation 
		(cf.\S \ref{sec:P(d)_solid_state})).
		 Again one replaces the basic micro theory $\textbf{m}$, as described in  \cite{Bishop_2005_patch} (quoted above), with an effective theory $\textbf{m'}$ that incorporates the approximations needed to make it work. In this case, one uses the fact that nuclei are far heavier than electrons to derive a new micro theory \textbf{m'} that breaks translational  symmetry and gives the required results. It is not the same as the theory you started with: it is the lower level effective theory you need.\\
	
Why does one introduce the effective theory $\textbf{m'}$ at the low level, rather than using the fundamental theory $\textbf{m}$  based only in Newton's Laws, QFT, Maxwell's equations, and so on? The answer is that if we use $\textbf{m}$, unless the coarse graining operation $\textbf{C}$ explicitly breaks the symmetry $\textbf{S}$, the correct emergent results cannot even in principle be deduced in a purely bottom up way. In notional terms,
\begin{equation}\label{eq:not_possible}
\{\textbf{S}(\textbf{m})=\textbf{m}, \,\, \textbf{C}(\textbf{m})=\textbf{M},\,\, \textbf{CS}=\textbf{SC}\} \,\,\,\Rightarrow\,\,\, \textbf{S}(\textbf{M}) = \textbf{M}
\end{equation}
contradicting the known macrolevel symmetry breaking of \textbf{M}. You cannot get \textbf{M} from \textbf{m}   
via any coarse graining \textbf{C} that does not explicitly break the symmetry \textbf{S}, so that $\textbf{CS} \neq \textbf{SC}.$  This could happen if  (cf. Eqn.(\ref{eq:downward})),  
 $\{\textbf{C} \rightarrow \textbf{C'} = \textbf{C'}(\textbf{a})\} \Rightarrow \textbf{SC'}\neq \textbf{C'S}$, but then you are specifically introducing a macro-determined term into this averaging process.

\begin{quote}
	\textbf{Conclusion: \textit{Strong emergence  of properties \textbf{P}(\textbf{d}) takes place when SSB(\textbf{M}) occurs: the symmetry broken higher level dynamics  cannot even in principle be obtained by coarse graining the fundamental theory $\textbf{m}$ because $\textbf{S}(\textbf{m})=\textbf{m}$. One has to coarse grain the effective micro theory m', of such a nature that  $\textbf{S}(\textbf{m'}) \neq \textbf{m'}$, to get the correct result \textbf{M} by coarse graining. 
		This covers the way emergence of properties occurs in condensed matter physics and in  quantum chemistry}.}
\end{quote} 
In short: you have to add a symmetry breaking term into the micro theory in order to get the correct macro theory, because it's not there in the fundamental physics; but you only  can work out what symmetry breaking term to add from your  knowledge of the correct macro theory \textbf{M}. You have to use variables defined by that higher level theory to get the symmetry broken effective micro theory \textbf{m'} which gives the correct macro result \textbf{M}. 

 

\paragraph{Change of symmetry changes weak to strong emergence} When \textbf{SSB}(M) takes place, weak emergence \textbf{E} occurs via  a \textit{First Order Phase Transition}.  

(\cite{Biney 1992}:1-2) describe it thus:
\begin{quote}
	``\textit{Under ordinary circumstances the phase transitions of $H_2O$, or the solidification of a molten metal, are `first-order' phase transitions ... that involve latent heat. When a material makes a first order phase transition from a high temperature phase to a low temperature phase, a non-zero quantity of heat, the latent heat, is given out as the material cools through an infinitesimally small temperature around the transition temperature $T_t$. This emission of heat at the transition tells us that the structure of the material is being radically altered reordered at $T_t$. ... Above the freezing point of water, there is no crystal lattice. Below the freezing point, the lattice is well defined even if not free of imperfections (`defects'). The transition from water to ordered ice is an all or nothing affair. Either the lattice is there and the vast majority of $H_2O$ molecules are comparatively tightly bound, or there is no lattice and the molecules are not optimally packed.
		''}
\end{quote}
In the light of the above discussion, its meaning is as follows:
\begin{quote}
	\textbf{First order phase transitions:} \textbf{\textit{As the temperature $T$ decreases through the transition temperature $T_t$,  Spontaneous Symmetry Breaking (\textbf{SSB}) takes place. Weak emergence \textbf{E} occurs as the material goes from a more symmetric to a less symmetric state, the binding energy of the less symmetric state being equal to the latent heat given out. Consequent on this symmetry change, the properties 
		\textbf{P} of the substance change from being weakly emergent to strongly emergent.}   }
\end{quote}
This is similar to what happens when molecules form out of atoms A key comment is the following: when this takes place the relevant Hilbert space of the problem changes completely. This is pointed out in (\cite{Anderson 1984}:126) in the case of molecules: ``\textit{A change of this sort puts the system, in the limit $N \rightarrow \infty$, into a wholly different, orthogonal Hilbert space, from which there is no easy continuous method of return''}. This is how properties \textbf{P} become strongly emergent when a crystal or molecular structure arises through weak emergence  \textbf{E} via
 \textbf{SSB}(\textbf{M}). \cite{Anderson 1984} emphasizes that SSB leads to singular points and thus a breakdown in continuity. 
 
 
\paragraph{Giving more details} The above  argument traces the key causal relations in what is going on. To give a more detailed proof, one would have \textit{inter alia} to parse the micro dynamics \textbf{m} into a reliable  and unchanging relation \textbf{L} (`the Laws of Physics') between initial conditions described by data $d$ and outcomes $o$, which is valid in some domain ${\cal D}$. That is, 
\begin{equation}\label{eq:Laws_of_physics}
\textbf{L}: d \in {\cal D} \rightarrow \textbf{L}[d] = o \in {\cal D}
\end{equation}
in a reliable way, whether \textbf{L} is an exact or statistical law.\footnote{Note that the laws of physics are not algorithms - it's Newton's Laws of Motion, not Newton's algorithm  - and they do not compute, see \cite{Binder Ellis 2016}.} 
One would have to represent the emergent macro dynamics \textbf{M} in a similar way to  (\ref{eq:Laws_of_physics}), and then carry out an analysis analogous to that above.

 I have not attempted this more complex project here 
 because I do not believe it would throw much light on what is going on: rather it would probably obscure the key  relations. I believe that the symbolism used above enables one to clearly understand the relevant causal links in a adequate way. 
  However I will develop Equation (\ref{eq:Laws_of_physics}) in important ways below: it will be modified to provide \textit{Effective Laws} in various contexts. Specifically, Equation (\ref{eq:Laws_of_physics_V}) applies in the case of dynamics with a potential shaped by a crystal structure, and  Equation (\ref{eq:Laws_of_biology}) when electron dynamics is controlled by molecular constraints. 
  
  It is important to relate (\ref{eq:Laws_of_physics}) to emergent levels of complexity \cite{Ellis_closure}. One can characterise an Effective Theory $\textbf{ET}_\textbf{L}$ valid at some level \textbf{L} as follows:
  \begin{quote}
	  \textit{An} \textbf{Effective Theory} $\textbf{ET}_\textbf{L}$ \textit{at
 an emergent  level \textbf{L} is a reliable relation between initial conditions described by effective variables $v_\textbf{L} \in \textbf{L}$ and outcomes $o_\textbf{L}\in \textbf{L}$:}
  	\begin{equation}\label{eq:effective_laws}
  	\textbf{ET}_\textbf{L}: v_\textbf{L} \in \textbf{L} \rightarrow \textbf{ET}_\textbf{L}[v_\textbf{L}] = o_\textbf{L} \in \textbf{L}
  	\end{equation}
 \textit{ in a reliable way, whether $\textbf{ET}_\textbf{L}$ is an exact or statistical law.
}\end{quote}   
 \paragraph{All well tested laws of physics are effective laws }
  The microphysics \textbf{m} that I have used as a base level is not in fact the fundamental level of physical laws. It itself arises out of lower levels - and we do not know what the bottom-most level is. It might be String  Theory/M Theory, but again it might not. No such Theory of Everything (TOE) is well defined, let alone experimentally verified. Consequently,  
 \begin{quote}
 	\textbf{The Laws of Physics are Effective Laws \textit{Despite the fact that they are effective, eternal, and unchanging, all the well established and tested laws on which physics and engineering are based are effective laws\footnote{They are effective theories in the sense of \cite{Castellani_02}.} of the form (\ref{eq:effective_laws}). This applies equally to Newton's Laws of motion, Newton's Law of Gravitation, Maxwell's equations, Einstein's gravitational field equations, the Schr\"{o}dinger equation,  Dirac's equation, and the Standard Model of Particle Physics. They all only hold in a restricted domain ${\cal D}$.}}
 \end{quote}
 This is discussed further in Section \ref{sec:equal_valid}. What this also implies is that a key aspect of determining any Effective Theory is to determine its Domain of Applicability ${\cal D}$.

\subsection{Downward Causation and Causal Completeness
} \label{sec:downward_diachronic_synchronic}
Equation (\ref{eq:downward}) is where the downward effect of the macro state on the micro state is explicitly represented, through the modification $\textbf{m} \rightarrow \textbf{m'}(\textbf{a})$. Thus Downward Causation occurs. The point is that \textbf{a}=\textbf{a(\textbf{M})} is  a variable that cannot even in principle be represented in terms of the microlevel variables occurring in \textbf{m}. The reason is that  \begin{quote}
	\textbf{\textit{All the variables occurring in  microphysics \textbf{m} respect the symmetry \textbf{S}, whereas \textbf{a}(\textbf{M}) - and so \textbf{m'} -  does not}}
\end{quote}
And as has been shown above, we must use \textbf{m'} as the micro level theory if we want to get the correct macro level results. The theory \textbf{m} is unable to do the job. The key effect underlying this physically in the case of solid state physics is the quantum  theory  wave(macro)-particle(micro)   duality discussed in Section \ref{sec:Phonons}, 
which leads to crucial lower level effective variables (quasiparticles) existing, without which the theory would not work.
\begin{quote}
	\textbf{Effective Dynamics and Downward Causation} \textbf{\textit{In order to derive the correct macrodynamics M, the microdynamics m and fundamental laws L must be replaced by effective microdynamics m' and effective laws L' respectively. The maps $\textbf{m} \rightarrow \textbf{m'}$, $\textbf{L} \rightarrow \textbf{L'}$  represent the effects of context on the functioning of physics at the micro level. That is, they represent downward causation.}}
\end{quote}

\paragraph{Causal completeness?}
It is claimed by many that such a downward causal influence is not possible because of the causal completeness of physics at the microlevel together with supervenience of the macro level \textbf{M} on the micro level \textbf{m}, see for example \cite{Handbook_of_Emergence} 
 which gives links to Kim who originated the argument.
There has to be something wrong with this claim, because of the remark just made:  the allegedly causally complete 
dynamics \textbf{m}  must be replaced by \textbf{m'} if you want your Effective Theory to accord with experiment. And \textbf{m'} can only be obtained by introducing variables not present in \textbf{m}.
The fact is that \textbf{m}  cannot do the job  (see (\ref{eq:not_possible})), so it must be causally incomplete.  
 
 Robert Bishop refutes the causal completeness claim in the context of fluid convection \cite{Bishop_2008_convection}. 
  In \cite{Ellis_2019_Godel}, I  give a refutation of the argument from causal completeness  based firstly in the issue of multiple realisability of a higher level state by lower level states (see Section \ref{sec_multiple}),  and secondly in the difference between synchronic and diachronic supervenience. 
Here, on the basis of the above results,  I will reverse the argument. 
\begin{quote}\label{Theorem_main}
	\textbf{Causal completeness of the microdynamics m. \textit{The microdynamics m arising purely out of the Laws of Nature L is incomplete in this sense: it cannot derive in a bottom-up way properties occurring in condensed matter physics, as just demonstrated. Causally complete   dynamics can however be attained by including downward effects in \textbf{m} that result in an effective lower level theory m' that is able to give the correct emergent results.}}
\end{quote}
This theme of interlevel causal completeness is developed in depth in \cite{Ellis_closure}. Here, I confirm this claim by investigating strong emergence \textbf{P} of properties  in condensed matter physics
, using the definitions given in Section \ref{sec:conceptual}. 

\subsection{
	Downward causation and Effective Laws}\label{sec:P(d)_solid_state}
 Consider properties of a crystal, such as its optical properties and electrical conductivity.  
An effective potential term at the lower level is provided by the symmetry-breaking lattice structure that is relatively immobile in comparison with electron motion, so it may be at first approximation be regarded as fixed (this is the \href{https://en.wikipedia.org/wiki/Born%E2%80%93Oppenheimer_approximation}{Born-Oppenheimer}
	 approximation that breaks translational invariance). The resulting potential $V(\textbf{q},\textbf{x})$ shapes the lower level dynamics. The symmetry breaking occurs via the fact that $\partial V(\textbf{q},\textbf{x})\partial \textbf{x} \neq 0$. In this case, equation (\ref{eq:Laws_of_physics}) is modified to give the effective laws \textbf{L'} as follows: $\textbf{L} \rightarrow \textbf{L'}(V)$ such that
\begin{eqnarray}\label{eq:Laws_of_physics_V}
\textbf{L'}(V):d \in {\cal D}\rightarrow 
\textbf{L'}(V)[d] = o \in {\cal D},\,\,
V = V(\textbf{q},\textbf{x}), \,\, 
\partial V(\textbf{q},\textbf{x})/\partial \textbf{x} \neq 0.
\end{eqnarray}
 Now  comes the essential point: because the inhomogeneity  leads to \href{https://en.wikipedia.org/wiki/Phonon}{lattice vibrations}, the potential  will be time dependent:\footnote{``\textit{The broken symmetry state is	not an eigenstate of the system Hamiltonian H (it breaks one of the symmetries of H) and so it is
 	not a stationary state.}'' \cite{Blundell 2019}.}
\begin{eqnarray}\label{eq:Laws_of_physics_time}
 \{\partial V/\partial x \neq 0 \Rightarrow 
\partial V/\partial t \neq 0\} \Rightarrow \textbf{L'}(V):d \in {\cal D} \rightarrow 
\textbf{L'}(V(\textbf{q,}\textbf{x},t))[d] = o \in {\cal D}.
\end{eqnarray}
The lower level physics is not causally closed: outcomes depend on the higher level lattice vibrations, as indicated by the time dependence of the potential $V(\textbf{q,}x,t)$. 
They are equivalent to the existence of quasi-particles such as phonons which shape causal outcomes at the lower level (Section \ref{sec:Phonons}).
\begin{quote}
	\textbf{
		Downward causation (solid state physics)}:  
\textit{		Symmetry breaking at the higher lead to a time dependent effective potential governing lower level dynamics. Equation (\ref{eq:Laws_of_physics}) is replaced by (\ref{eq:Laws_of_physics_time}) and outcomes are determined by the time dependence of the potential}. 
\end{quote}
This supports the analysis above and in \cite{Ellis_2019_Godel} arguing for strong emergence when diachronic supervenience take place. 
However much more than that, (\ref{eq:Laws_of_physics_time}) can be extended to represent downward action of higher level variables controlling lower level causal effects, as I now illustrate in the specific case of a transistor in a digital computer. Interlevel causal closure then involves many higher levels \cite{Ellis_closure}.
\subsection{Time dependent potentials: Transistors in computers}\label{sec:details}
A specific context where time dependent potentials control lower level dynamics is  the case of transistors in a digital computer, where algorithms encoded in a high level language  
chain down via compilers to generate sequences of bits that control electron flows in transistors \cite{Tannenbaum}. This is discussed in depth in \cite{Ellis_Drossel_computers}.
I will reproduce here the key part of that argument, showing how time dependent potentials generated in that way control the electron  dynamics.
 
\paragraph{The Hamiltonian} description at the electron/ion level is  (\cite{Philips}:16):
\begin{eqnarray}
H &=& 
-\sum _i\frac  {\hbar ^2}{2M_i}\nabla _{{\mathbf  {R}}_i}^2
-\sum _{i}\frac {\hbar ^{2}}{2m_e}\nabla _{\mathbf {r} _{i}}^{2}
+ \sum _i\sum _{j>i}\frac  {Z_{i}Z_{j}e^{2}}{4\pi \epsilon _{0}\left|{\mathbf  {R}}_{i}-{\mathbf  {R}}_{j}\right|} \nonumber\\
&& -\sum _{i}\sum _{j}{\frac  {Z_{i}e^{2}}{4\pi \epsilon _{0}\left|{\mathbf  {R}}_{i}-{\mathbf  {r}}_{j}\right|}}
+ \sum _{i}\sum _{{j>i}}{\frac  {e^{2}}{4\pi \epsilon _{0}\left|{\mathbf  {r}}_{i}-{\mathbf  {r}}_{j}\right|}}\label{eq:Hamiltonian}
\end{eqnarray}
To derive from this the effective Hamiltonian encoding the  lattice structure and electron charge distribution in a transistor, we must make a series of approximations. 

A:  First, electrons are characterised as either conduction band electrons (essentially unbound and so free to move) or valence band electrons (closely bound to ions and so localised). This represents  the Hamiltonian (\ref{eq:Hamiltonian}) in the form 
\begin{equation}
H = T_i + T_e + V_{ii} + V_{ee} + V_{ei} + E_{core}
\end{equation}
Note that this is where knowledge of the higher (crystal) level dynamics is injected downwards into the representation of the lower level (ion/electron) dynamics (one can't make that splitting without that knowledge). 

B: Now one uses the Born-Oppenheimer (adiabatic) approximation, thereby altering the Hamiltonian used. Thus this represents the transition (\ref{eq:downward}) discussed in Section \ref{sec:SSB(M)}.
This approximation assumes that the electrons are in equilibrium with the locations of the ions (\cite{schwablqm},Ch.15).
The wave function is factorized into two parts: the  electron part $\Psi_{e}(\textbf{r},\textbf{R})$ for given positions of the ions, and the ion part $\Phi(\textbf{R})$,
\begin{eqnarray}
\Psi(r,\textbf{R}) =  \Phi(\textbf{R}) \Psi_{e}(\textbf{r},\textbf{R}) \, .
\end{eqnarray}
This results in an electron equation 
\begin{eqnarray} \label{electronequation}
(T_e+V_{ee}+V_{ei}) \Psi_{e}(\textbf{r},\textbf{R}) = E_{e}(\textbf{R})\Psi_{e}(\textbf{r,R})
\end{eqnarray}
and an ion equation  
\begin{eqnarray}\label{ionequation}
(T_i+V_{ii}+E_{core}+E_{e}(\textbf{R})) \Phi(\textbf{R}) =  E\Phi(\textbf{R})  \, .
\end{eqnarray}

C: Using the electron equation (\ref{electronequation}), one  obtains the electronic band structure. This is based on a picture of non-interacting electrons, so the interaction term $V_{ee}$ must be dropped and the electron Hamiltonian becomes a sum of one-particle Hamiltonians. The symmetry breaking contextual term determining electron outcomes is $V_{ei}$.  In the context of a specific  periodic lattice, the solutions are Bloch waves. These give the energy bands $E_n(\textbf{k})$, which are the lattice analog of free particle motion (\cite{Phillips2012}:22).

D: Electron-lattice interactions occur via phonons (\cite{Simon 2013}:82-84, 90-95). Using the ion equation (\ref{ionequation}) and further approximations (ions are localized, harmonic expansion of  energy around equilibrium ion positions), one can derive the phonon modes of the ions. The symmetry breaking term here is $V_{ii}$.  The dispersion relations of phonons are determined by the normal modes of the resulting harmonic model.

E: In order to  explicitly model electron-phonon interactions, a quantum field theoretical formalism is needed based on  creation and annihilation operators for electrons and phonons (\cite{Solyom 2009}, Ch.23). The dispersion relations of electrons and phonons obtained in this way determine the electron and phonon propagators, however evaluating the interaction term needs a separate determination of the cross section for scattering of electrons via emission or absorption of a phonon. 

F: Doping with impurities adds electrons to the conduction band (`donor' n-doping) or holes to the valence band (`receptor' p-doping) (\cite{Simon 2013}:187-194).
The junctions between n-doped and p-doped regions are the key feature of transistors (\cite{Mellis}:14-20; \cite{Simon 2013}:199-203). Electron diffusion at such junctions leads to depletion regions, stabilized by induced electric fields (\cite{Mellis}: Fig.1.9). 
Transistors are created by suitably shaped such junctions (\cite{Simon 2013}:203-205).  

\paragraph{Electric voltage effects} In order to model the steady state situation when the transistor is in either a conducting state or not, depending on whether the applied voltage is above a threshold or not, electric field effects must be added. Thus the gate  voltage  $V(t)$ must be added to the model. 
This leads to an added potential energy term  in the Hamiltonian of the electrons:
\begin{equation}\label{HV} 
H_V(t)=\sum_i e V(\textbf{r}_i,t)\end{equation}
where the higher level variable $V(t)$ determines the lower  level variables $V(\textbf{r}_i,t)$ in a downward way. The electrons  reach an equilibrium  where the electrical field created by the modified charge distribution cancels the electrical field due to the gate voltage. To calculate this equilibrium and so determine $H_V(t)$, a self-consistent calculation  is needed, based on the charge density due to doping, the gate potential, and the thermal excitation. 

This alters the band structure according to the bias  voltage applied,  and thereby either creates a conduction channel by changing the depletion zone, so current flows, or not,   so no current flows (\cite{Mellis}:19),  (\cite{Simon 2013}:Fig.18.6).  

\begin{quote}
	\textbf{In summary}: \textit{This explicit example shows 
		\textbf{(a)} how the basic Hamiltonian (\ref{eq:Hamiltonian}) representing the microdynamics \textbf{m} gets altered, in real world applications, to give the effective dynamics \textbf{m'} as in (\ref{eq:downward}), affected by the higher level state; \textbf{(b)} the time dependent potential term (\ref{HV}) determines lower level dynamical outcomes to give an extension of (\ref{eq:Laws_of_physics_time}) where now the time dependence of $H_V(t)$ is determined by the logic of algorithms; \textbf{(c)} this is a clear example of downward causation enabled by the modular hierarchical structure of the computer \cite{Tannenbaum}. The word `causation' is justified in both counterfactual and experimental terms: a different algorithm results in different electron flows. \textbf{(d)} It is a convincing case of strong emergence in Chalmer's sense \cite{Chalmers_2000} because the physics summarised in (\ref{eq:Hamiltonian}) cannot in principle by itself lead to deduction of the kind of logical branching characterising the functioning of the emergent computational dynamics.\footnote{This is in parallel to the biological case, see Section \ref{Bio_variables} and equation (\ref{eq:choice}) below.}}  
\end{quote}
  Algorithms produce real world outcomes by this chain of downward  causation.

\subsection{Multiple Realisability}\label{sec_multiple}
A key point is that multiple realisability plays a fundamental  role in strong emergence \cite{Menzies 2003}. Any particular higher level state can be realised in a multiplicity of ways in terms of lower level states. In engineering or biological cases, a high level need determines the high level function and thus a high level structure  that fulfills it. This higher structure is realised by suitable lower level structures,  but there are billions of ways this can happen.
It does not matter which of the equivalence class of lower level realisations is used to fulfill the higher level need,  as long as it is indeed fulfilled. Consequently you cannot even express the dynamics driving what is happening  in a sensible way at a lower level. \\

Consider for example the statements \textit{The piston is moving because hot gas on one side is driving it } and \textit{A mouse is growing because the cells that make up its body are dividing}. They cannot sensibly be described at any lower level not just because of the billions of lower level particles involved in each case, but because  \textit{there are so many billions of different ways this could happen at the lower level}, this
cannot be expressed sensibly at the proton and electron level. 
The point is the huge different numbers of ways combinations of lower level entities can represent a single higher level variable. Any one of the entire equivalence class at the lower level will do. Thus it is not the individual variables at the lower level that are the key to what is going on: it is the equivalence class to which they belong. But that whole equivalence class can be describer by a single variable at the macro level, so that is the real effective variable in the dynamics that is going on. This is a kind of interlevel duality:
\begin{equation}\label{eq:Equiva_class}
\{v_\textbf{L} \in \textbf{L} \}\Leftrightarrow \{v_\textbf{i}: v_\textbf{i} \in E_{\textbf{L-1}}(v_\textbf{L-1})\in \textbf{(L-1)}\} 
\end{equation}
where $E_{\textbf{L-1}}(v_\textbf{L-1})$ is the equivalence class of variables $v_\textbf{L-1}$ at Level $\textbf{L-1}$ corresponding to the one variable $v_\textbf{L}$ at Level \textbf{L.} The effective law $\textbf{EF}_\textbf{L}$ at Level \textbf{L} for the (possibly vectorial or matrix) variables $v_\textbf{L}$ at that level is equivalent to a law for an entire equivalence class $E_{\textbf{L-1}}(v_\textbf{L-1})$ of variables at Level \textbf{L-1}. It does not translate into an Effective Law for natural variables $v_{\textbf{L-1}}$ \textit{per se} at Level \textbf{L-1}. The importance of multiple realisability is discussed in \cite{Ellis_2019_Godel} and \cite{Bishop_Ellis}. The latter paper also discusses the underlying group theory relations, dealing in particular with the emergence of molecular structure. 

\begin{quote}
	\textbf{Essentially higher level variables and dynamics}
\textit{The higher level concepts are indispensible when multiple realisability occurs, firstly because they define the space of data $d_\textbf{L}$ relevant at Level \textbf{L}, and secondly because of  (\ref{eq:Equiva_class}), variables in this space  cannot  be represented  as natural kinds at the lower level. Effective Laws $\textbf{EF}_\textbf{L}$ at level \textbf{L} can only be expressed at level \textbf{L-1} in terms of an entire equivalence class at that level. One can only define that equivalence class by using  concepts defined at level \textbf{L}.}
\end{quote}
To inject reality into this fact, remember that the equivalence class at the lower level is typically characterised by Avagadro's number. 

Related to this, one should note that the dynamic or even static properties \textbf{P}  of a macro system are the result of much faster dynamics at the microlevel. Electrons and molecules are always in motion, even ions in a crystal lattice are vibrating.  Properties such as strength and ductility and brittleness are due to atoms moving past each other (as they are in metals), and this depends on dislocations, which are dynamic phenomena at the micro level, by which metallurgists mean microns rather than nanometres.
\section{Emergence and Life}\label{sec:emerge_life}
 The argument for strong emergence is quite different in the cases of biological and physical systems. 
  In this section I first comment on purpose and life (Section \ref{sec:purpose}). Then I argue that the processes of Natural Selection are strongly emergent (Section \ref{sec:natural_select_emerge}). 
   I   use an argument based in the existence of time dependent constrains to show that molecular biology dynamics is strongly emergent (Section \ref{sec:biology_diachronic}). In Section \ref{Bio_variables}, I comment on whether there could be a dictionary from physics to biology. Putting this together,  I argue that strong emergence certainly   occurs in biology  (Section \ref{sec:Strong_Emerge_Biol}).  
\subsection{Purpose and Life}\label{sec:purpose}
 Do biological organisms have purpose? It is strongly argued by Nobel Prize winning biologist \href{https://www.nobelprize.org/prizes/medicine/2001/hartwell/biographical/}{Leland Hartwell} and colleagues that this is indeed the case, in \cite{Hartwell_et_al_1999}: 
\begin{quote}\textit{``Although living systems obey the laws of physics and chemistry, the notion of function or purpose differentiates biology from other natural sciences. Organisms exist to reproduce, whereas, outside religious belief, rocks and stars have no purpose. Selection for function has produced the living cell, with a unique set of properties that distinguish it from inanimate systems of interacting molecules. Cells exist far from thermal equilibrium by harvesting energy from their environment. They are composed of thousands of different types of molecule. They contain information for their survival and reproduction, in the form of their DNA. Their interactions with the environment depend in a byzantine fashion on this information, and the information and the machinery that interprets it are replicated by reproducing the cell.}''
\end{quote}
Even proteins have functions \cite{Petsko_Ringe_2009}. One can claim that the interaction between structure and function whereby such function and purpose is realised is central to biology \cite{Campbell and Reece 2008}. 

  A central point as regards the nature of emergence is that words that correctly describe causation at higher levels simply do not apply at lower levels. It is not a point of confusion, it's a central aspect of emergence.  As stated by  \cite{Anderson_72}, ``\textit{each level can require a whole new conceptual structure}''. Just so; and that means new terminology and new variables $v_\textbf{L}$ at higher levels \textbf{L}. To make this clear. I give a number of examples where this occurs in Section \ref{sec:natural_select_emerge}. The issue of whether there could be a dictionary deriving these variables from the underlying physics is considered in Section \ref{Bio_variables}.
 In the case of biology, unless the higher levels include the concept  of function, they will miss the essence of what is going on, as pointed out by \cite{Hartwell_et_al_1999}. This leads to the key role of information flows \cite{Davies 2019} and logic \cite{Nurse 2008} \cite{Ellis_Kopel_2019}  in the processes of life. This crucially distinguishes biology from physics. 

But how does this all emerge from physics? Through symmetry breaking, as in the physics case discussed above, together with time dependent constraints and the extraordinary dependence of molecular biology on the conformational shape of biomolecules \cite{Ellis_Kopel_2019}. Wonderful examples are given in \cite{Karplus_2014}. This all came into being via  Darwinian processes of natural selection \cite{Mayr 2001}:  one of the best-established processes in biology, which is strongly emergent.  

\subsection{Why Natural Selection is Strongly Emergent}\label{sec:natural_select_emerge}
 Developmental processes leading to the existence of proteins, and hence cells and bodies,  take place by reading the genome via processes controlled by gene regulatory networks. This is emergence  \textbf{E} of macro structure from microstructure. But it is based in the current existence of the genome and developmental systems. The genome has famously been shaped by processes of \href{https://en.wikipedia.org/wiki/Natural_selection}{natural selection} \cite{Mayr 2001} taking place over evolutionary timescales. The key element is 
 \begin{quote}
 	\textit{\textbf{Evolutionary Emergence}: Genotypic mutations that provide some phenotypic
 		advantages given some environmental conditions enable the species at stake to survive (relative to
 		other species).}
 	 \end{quote}
When this occurs, the outcomes cannot even in principle be deduced from the underlying microphysics, because both of the random aspect of the processes of natural selection, and their top-down nature \cite{Campbell_1974}. Apart from random drift, which is a significant effect in evolutionary history, life today is determined by adaptation to a variety of physical, ecological, and social contexts in the past. There was randomness in the process due to mutation, recombination, and drift at the genetic level, which was influenced by molecular randomness as well as by quantum events that cannot be predicted in principle (emission of cosmic rays that caused genetic mutations)  so evolution from initial data as determined purely by the underlying physics cannot in principle determine unique outcomes (see the discussion in Section \ref{sec:determine}). Rather time-dependent downward influences took place \cite{Campbell_1974} due to the time varying niches available  for adaption, which determined what superior variation was selected. According to Ernst Mayr \cite{Mayr 2001},  
\begin{quote}
	\textit{``Owing to the two step nature of natural selection, evolution is the result of both chance and necessity. There is indeed a great deal of randomness (`chance') in evolution, particularly in the production of genetic variation, but the second step of natural selection, whether selection or elimination, is an anti-chance process.'' \footnote{However note that the issue of randomness is complex and subtle; see for example \cite{Wagner_2012_random}. }
}\end{quote} 
He then gives the eye as an example, which enhances relative survival probabilities because of the information processing and consequent contextually-dependent predictions of outcomes it allows. Neither step is predictable from physics, even in principle, because physics does not have the conceptual resources available to undertake the discussion. They are essentially biological in nature, and hence strongly emergent: the causal relations governing what happens (as described by Mayr) occur at emergent biological levels through interactions between biomolecules such as DNA, RNA, and proteins, determined by their specific molecular structures. Thus  
we have here \textbf{SB}(\textbf{NS}): the specific nature of the broken symmetries occurring in biomolecules today \cite{Petsko_Ringe_2009}, and so shaping current biological outcomes,  has been determined by Darwinian processes of natural selection (\textbf{NS}) during our past evolutionary history leading to existence of these molecules. 

To strengthen the point, I will now consider cases of strong emergence 
 in biology due to such selection   
at the molecular level. 
Biological function at the molecular level is determined by the physical nature of biomolecules, with \href{https://en.wikipedia.org/wiki/Protein}{proteins} 
being the workhorses of biology. Due to the nature of the underlying physical laws and the values of the constants occurring in the Laws of Physics (\ref{eq:Laws_of_physics}), only certain proteins are possible in physics terms. Thus there  is an abstract possibility space for the existence of proteins, as discussed by  Andreas Wagner \cite{Wagner 2014}. It is of enormous dimension, as he emphasizes.
The proteins that actually exist on Earth, out of all the possible ones that might have existed but have not been selected for, have a very complex structure that is determined by the functions they perform \cite{Petsko_Ringe_2009} (note the use of the word ``function''). They were selected by Darwinian processes \cite{Mayr 2001} so as to have a structure that will adequately perform a needed function. 

An example is the %
Arctic Cod. 
Wagner discusses selection for proteins in Chapter 4 of his book (\cite{Wagner 2014}:107). The Arctic cod lives within 6 degrees of the North Pole in sea water whose temperature regularly drops below 0 degrees Celsius. This ought to destroy the cells in the fish, as ice crystals should form within  the cells and tear them apart.  To deal with this problem, the arctic cod survives by producing antifreeze proteins that lower the temperature of its body fluids. What has been selected for is not just the genome that codes for  the protein that will do the job, but the entire developmental system that produces the protein, involving gene regulatory cascades   (\cite{Wagner 2014}:Chapter 5). This is a case of downward causation from the environmental level (the cold nature of the sea water) to details of the DNA sequence that does the job \cite{Ellis_2016}. Another example is the 
precisely patterned distribution of a particular family of proteins in squid eyes that leads to a graded refractive index in its lenses \cite{lenses}. A  parabolic relationship between lens radius and refractive index allows spherical lenses to avoid spherical aberration \cite{Sweeney_2007_squid}. This physiological advantage results in evolutionary development of specific proteins  
that are then precisely positioned to attain this effect.  
Thus the need for clear vision in an underwater environment lead to selection of a genotype that produces this outcome via squid  developmental systems.

 The point of these examples is that causation is real at these biological levels. Tee underlying physics enables this all to happen, but does not determine the specific outcomes: it cannot do so as a matter of principle. 
  The emergent relations are determined by biological needs, which is
why it is essential to use language appropriate to the higher level in order to get a model that can provide a  reductionist understanding of mechanisms. 

Similar arguments apply at the cellular level, for example in determining the way neurons have specific voltage gated ion channels imbedded in axon membranes that underlie the Hodgkin-Huxley equations for action potential propagation \cite{Hodgkin_Huxley}. As discussed by \cite{Scot 1999}, the constants in those equations cannot in principle be determined by the underlying physics as they are determined by their physiological functioning. If they were determined in a bottom up way, those constants would be determined by the fundamental physics constants. This proves this is strong, not weak, emergence. 



\paragraph{The selection relation}
 When natural selection occurs, there is a phenotype level  \textbf{LP} where phenotypic advantage leads to enhanced relative survival rates after genotype mutations occur. The genotype level \textbf{LG} upwardly produces outcomes at the phenotype level \textbf{LP} through developmental processes, in a contextually dependent way.  
Selective choice of advantageous phenotypes $v_{\textbf{LP}}$ at the higher level \textbf{LP} (or at least ones that are not seriously disadvantageous) chains down to select preferred genotypes $v_{\textbf{LG}}$ at the lower level \textbf{LG}. The effective theory  $\textbf{ET}_\textbf{LG}$ at the genotypic level (reading of genes to produce proteins via developmental systems) is altered  because the set of variables $\{v_\textbf{LG}\}$ at level \textbf{LG} has been changed by the selection process $\textbf{S}_\textbf{LG}({v}_\textbf{LP})$ to give a new set of variables $\{v'_\textbf{LG}\}$ at that level that will produce the preferred outcomes $\{v_{\textbf{LP}}\}$ at the phenotype level: 
\begin{equation}\label{eq:adaptive_select}
\textbf{S}_\textbf{LG}({v}_\textbf{LP}) :
\{v_\textbf{LG}\} \in \textbf{LG} \rightarrow \{v'_{\,\,\,\textbf{LG}}\} = \textbf{S}_\textbf{LG}({ v}_\textbf{LP})[v_\textbf{LG}]    \in \textbf{LG}
\end{equation}
This is thus a case of  \textit{Downward Causation by Adaptive  Selection}, deleting lower level elements. 
This enables alteration of structures and functions  at level \textbf{LG} so as to meet new challenges at level \textbf{LI}, in an interlevel closed causal loop. This is of course a highly simplified view of a complex process: in fact multiple phenotype levels are involved, and neutral evolution can occur. The fundamental point is unchanged.

	\subsection{Emergence of Properties \textbf{P} in  Biology}\label{sec:biology_diachronic}
	
	Downwards effects in a biological system occur 
	 because of physiological processes, see  \href{http://www.musicoflife.website/}{\textit{The Music of Life}} by Denis Noble,  and \cite{Noble2008_Music}, \cite{Noble 2012}. These processes are mediated at the molecular level by developmental systems \cite{Oyama Griffiths Cycles} operating through 
	gene regulator networks \cite{Wagner 2014} and cell signalling networks \cite{Berridge}, guided by higher level physiological needs. They reach down to the underlying physical levels via time dependent constraints on the lower level dynamics \cite{Ellis_Kopel_2019}, which is why the lower level physics \textit{per se} is not causally complete (cf. \ref{sec:downward_diachronic_synchronic}). The \textit{set of interactions} between elements at that level is uniquely characterised by the laws of physics \textbf{L}, but their \textit{outcomes} are determined by the biological context in which they operate, leading to effective laws \textbf{L'}. Equation (\ref{eq:Laws_of_physics}) should, in parallel with (\ref{eq:Laws_of_physics_V}),  be modified to read
	\begin{eqnarray}\label{eq:Laws_of_biology}
	\textbf{L'}:d \in {\cal D}\rightarrow 
	\textbf{L'}(C)[d] = o \in {\cal D},\,\,C = C(\textbf{a},\textbf{p},\textbf{q}), \,\, \partial C(\textbf{a},\textbf{p},\textbf{q})/\partial \textbf{a} \neq 0
	\end{eqnarray}
	where $C = C(\textbf{a},\textbf{p},\textbf{q})$ are constraints on microlevel coordinates $\textbf{p}$ and momenta $\textbf{q}$ that are dependent on biological variable \textbf{\textbf{a}} representing conditions at a higher level. Now comes the essential point (cf.(\ref{eq:Laws_of_physics_time})): because \textbf{a} is a biological variable, it will be time dependent:
	\begin{eqnarray}\label{eq:Laws_of_biology_time}
	\textbf{L'}:d \in {\cal D} \rightarrow 
	\textbf{L'}(C(\textbf{a}(t),\textbf{p},\textbf{q}))[d] = o \in {\cal D},\,\, \partial \textbf{a}/\partial t \neq 0 \Rightarrow 
	\partial C/\partial t \neq 0.
	\end{eqnarray}
	The lower level physics is not causally closed: outcomes depend on the time variation of the constraint C(t,\textbf{p},\textbf{q}).\footnote{A simple analogous model of how this effect works is a pendulum with time dependent length, see the Appendix of \cite{Ellis_Kopel_2019}.} This accords with Juarrero's characterisation of constraints as causes \cite{Juarrero 2002}.
	
	Examples are the voltage across a \href{https://en.wikipedia.org/wiki/Voltage-gated_ion_channel}{voltage gated ion channel} \cite{Ellis_Kopel_2019}, and the presence or not of a ligand bound to a \href{https://en.wikipedia.org/wiki/Ligand-gated_ion_channel}{ligand gated ion channel} \cite{Berridge}, which both alter molecular conformations and so change constraints on ion flows. To model this in detail requires quantum chemistry methods well adapted to this biological context, such as those used in  \cite{Karplus_2014}. Putting this in a real biological context such as 
	determination of    \href{https://en.wikipedia.org/wiki/Heart#Heart_rate}{heart rate} , pacemaker activity of the heart is via cells in the sinoatrial node that create an action potential and so alter ion channel outcomes. This pacemaking circuit is an integrative characteristic of the system as a whole \cite{Fink_and_Noble} -  that is, it is an essentially higher level variable - that acts down to this molecular level.
	
	In this way downward causation  takes place in biology and enables strong diachronic emergence of properties \textbf{P} in biology:
	\begin{quote}
		\textbf{Strong Emergence of Properties (biology):  	 \textit{Emergence of Properties \textbf{P} in biology is strong emergence because time dependent constraints  alter lower level dynamical outcomes in accord with  higher biological needs. Equation (\ref{eq:Laws_of_physics}) is replaced by (\ref{eq:Laws_of_biology_time}).} }
	\end{quote}

\subsection{A Dictionary from Physics to Biology?} \label{Bio_variables}
Along with Ernst Mayr \cite{Mayr 2001}, it  is my view that physics does not begin to have the resources to deal with this, \textit{inter alia} because the concepts of being alive or dead cannot be captured in physics terms, where only interactions between particles are characterised, so `survival'- central to evolutionary theory - is not a concept physics can deal with. Whether an animal is alive of dead is a biological issue, related to interlevel causal closure including  biological variables  \cite{Ellis_closure}. Life has crucial characteristics of its own \cite{Campbell and Reece 2008} that are orthogonal to concepts encompassed by physics.

Nevertheless I have been challenged by a referee as followings: might there be  a dictionary relating physical variables to  biological variables, thereby undermining my claim?  

While there are indeed dictionaries relating aspects of physics to key aspects of life, for example relating the physics of thermodynamics to use of energy in biology, and relating physics scaling laws to  biological scaling laws, such dictionaries fail to capture 
 the aspect of purpose or function focused on by Hartwell et al \cite{Hartwell_et_al_1999} and the use of information emphasized by \cite{Nurse 2008} and \cite{Davies 2019}.  Regarding energy use, life is famously an open thermodynamic system  \cite{Peacock 1989} \cite{Hoffman} where  metabolism is taking place, but this feature \textit{per se} does not characterise life, because even a candle flame is open in that sense, and it is certainly not alive.    
Indeed it is notoriously hard to state just what such conditions are. Various characterisations of \href{https://en.wikipedia.org/wiki/Life#:~:text=Life%20is%20a%20characteristic%20that,and%20are%20classified%20as%20inanimate.}
	{life} have been given, such as 
	\begin{itemize}
		\item Life involves Homeostasis, Organization, Metabolism, Growth, Adaptation, Response to Stimuli, and Reproduction with variation
		\item ``Life is a self-sustaining chemical system capable of Darwinian evolution.''
		\item ``Living things are self-organizing and autopoietic (self-producing)''
		\item Life is characterised by causal closure \cite{Mossio 2013},
		\item Life is characterised by function and purpose \cite{Hartwell_et_al_1999}
		\item Life is determined by information use \cite{Nurse 2008} \cite{Davies 2019} and associated information processing via logical operations \cite{Hoffman} \cite{Ellis_Kopel_2019} 
	\end{itemize}
and so on. None of these characterisations resemble any aspect of physics as commonly understood, either at the fundamental level of particles, forces and fields, or at emergent physics levels, except perhaps in regard to energy use and minimisation.   I will pursue just two aspects here: life as characterised in energy terms, where one might claim there are such dictionaries, and life as characterised in information processing terms, where it is implausible there is such a dictionary. 

\paragraph{Life as characterised in energy terms?}
Life certainly depends on energy, so some physicists have attempted to characterise life as being essentially described by energy or thermodynamic relations. I will comment on three different aspects here. 

The most sustained effort is that by Karl Friston \cite{Friston 2012}, developed into a major theory  (``The Free Energy Principle'') \textit{inter alia} encompassing the brain \cite{Friston 201O}. However this is not derived in a bottom up way from the underlying physics: rather it is an effective higher level theory constructed in analogy with  an important aspect of physics. It is about  free energy in the information theory sense, not literally in the thermodynamic sense, thus it is in effect an analogue of how energy functions in the purely physical domain. It is unrelated to the emergence of biology from the underlying lower level phyics. 

Second, one can suggest \cite{LM} that the ``purpose'' of some biological organism, seen from
our limited point of view, is procreation and the continuation of its species, but at the
same time is equivalent to the minimization of energy in some very complex phase space. Indeed so.  This is the concept of a \href{https://en.wikipedia.org/wiki/Fitness_landscape}{Fitness Landscape} as proposed by Sewell Right \cite{Wright 1932}, and developed by many others (see for example \cite{McGhee 2006}).  But this does nothing to further the reductionist project. It is an \textit{Effective Theory} in the sense explained in \cite{Castellani_02} (quoted in \S \ref{sec:EffectiveTheories}), but has nothing to do with weak emergence for example via an Effective Field Theory  relating microphysics to emergent phenomena by a power series expansions in terms of a physical parameter. A Fitness Landscape cannot be derived even in principle from the underlying physics by any deductive process that purely involves physics  concepts, precisely because it depends on the concept `fitness', and it does not map from the level of protons and electrons to the world of biological reality. It can however be developed as a productive  effective theory if one introduces relevant contextually dependent higher level biological concepts. 

Third, the way energy in the physical sense relates to biological activity is through the joint effects of the cardiovascular and digestive systems. These are macroscopic physiological systems, with  extremely well established emergent laws of behaviour that underlie medical practice in these respective areas. They interact in an integrated with way with immensely complex metabolic networks at the cellular level, where logic of the form (\ref{eq:choice}) determines what happens.  An example is the Citric Acid Cycle (\textbf{Figure 1}). This is without doubt causally effective at the cellular level and is irreducible to any lower level because of the closure of the major loop in the interaction diagram (it does not follow from the interactions of protons and electrons \textit{per se}). Notice also that this diagram establishes that a casual relation exists in Pearle's sense \cite{Pearl 2009} \cite{Pearl_Why}.

\begin{figure}[h]
	\centering
	\includegraphics[width=0.7\linewidth]{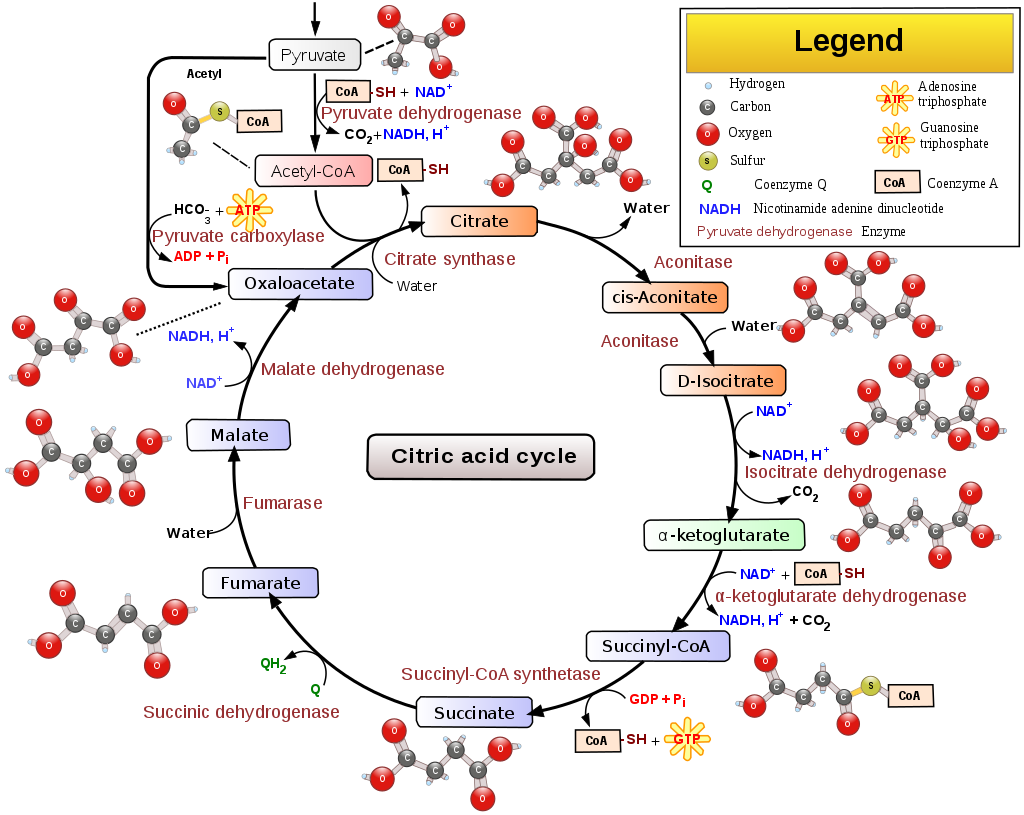}
	\caption{\textbf{Citric Acid Cycle} (Source: Wikipedia)}
	\label{fig:citricacidcycle}
\end{figure}

\noindent Metabolism at this level 
is driven by the dynamics at the emergent physiological system level  \cite{Noble 2012}, which characterizes relations between macro variables such as cystolic blood pressure and heart rate \cite{Fink_and_Noble}, and reaches down to control energy resources at the cellular and molecular levels \cite{Noble 2012}. These are inherently higher level variables, related to the detailed physiological structure and functioning of the heart.  These effective higher level relations \cite{Physiology_human} are also experimentally verified to high precision. This is a far cry from a process of energy or entropy minimisation. It is a highly coordinated topologically closed set of far from equilibrium interactions. Indeed it is controlled by information and logic, which I now discuss.

\paragraph{Life as characterised by information and logic} A contrasting view 
 is that Information is key in biology \cite{Nurse 2008} \cite{Davies 2019}, and the  
associated logic  cannot emerge out of non-information because information conveys meaning only in a specific higher level context. At each level biology operates according to a branching logic of the following form (\cite{Hoffman}:151-152) \cite{Ellis_Kopel_2019}: a signal $X$ is evaluated by a Truth Function $T(X)$ in a context ${\cal C}$, as follows 
\begin{equation}\label{eq:choice}
\textrm{GIVEN context } {\cal C,} \textrm{ IF  }  T(X) \textrm{ THEN }  O  \textrm{ ELSE }  O2 .
\end{equation} 
where $O1$ and $O2$ are alternative dynamical paths. This is for example the way all cell signalling works \cite{Berridge} and so underlies metabolic networks and gene regulatory networks. This kind of logic is implemented by the conformational shapes of the relevant biomolecules and their interactions with each other, allowed by the underlying physics but not reducible to it because physical laws \textit{per se} do not entail logical outcomes such as   (\ref{eq:choice}), and neither does specifying the detailed distribution of molecules (\ref{eq:L1}). This dynamics only emerges in the context of interactions at the cell level \cite{Hofmeyer 2017 Anticipation} \cite{Hofmeyer 2018 Constructors}, allowing interactions of the form (\ref{eq:choice}) to  occur and be causally effective, reaching down to alter outcomes at Level \textbf{L1}. 
At the cellular level, the important networks are cell signalling networks \cite{Berridge}, metabolic networks \cite{Wagner 2014}, and gene regulatory networks \cite{Carroll 2008}. They each have their own logic, irreducible to lower levels precisely because they are networks characterised by their topology as well as interaction strengths and timescales.  Because of multiple realisability, they can be realised in many billions of ways at the particle level. 

Now one can perhaps claim (\ref{eq:choice}) is somewhat like a phase transition (Section \ref{sec:SSB(M)}), associated with a change of symmetry. Indeed it is, but there are two key remarks here. First 
 a phase transition is an example of downward causation associated with strong emergence, because a macro level change of conditions (temperature $T$ decreasing) causes the micro level change of symmetry (\cite{Ellis_2016}:139) \cite{Green and Batterman 2020}. 
 Second, 
  in the case of biology, (\ref{eq:choice})  is associated with signalling \cite{Berridge}, while in the case of phase changes in physics, that element is absent.

However one point remains: Quantum Computing contains logical branching such as is characterised by (\ref{eq:choice}) when \href{https://en.wikipedia.org/wiki/Qubit#:~:text=In%20quantum%20computing%2C%20a%20qubit,with%20a%20two%2Dstate%20device.}
	{Qbits} are manipulated. Why does this not disprove what I have just claimed, as Qbits are basic physical phenomena, in effect occurring in any\href{https://en.wikipedia.org/wiki/Two-state_quantum_system}{two-state quantum system?} Is that not an example of a potential  dictionary from physics to biology?	
	The response is that Qbits can only be read out to produce branching logic such as (\ref{eq:choice}) in contexts that are extremely carefully engineered to make this outcome possible: in short, in the context of engineering systems. And such systems are strongly emergent \cite{Ellis_closure}. The physics by itself  cannot produce that kind of branching logic outside such a context. 
	 One can indeed produce dictionaries relating properties of such engineering systems to life and the brain, indeed that is what simulations are about. This ends up ultimately in the highly contested area of artificial intelligence. 
	   It is not a mapping between physics \textit{per se}, as characterised by interactions between particles and fields, and biology; it is between engineering systems and biology, an  entirely different proposition. \\
	
\noindent	Thus it is my view that
	\begin{quote}
		\textbf{Relating physics to life} \textbf{\textit{There can be dictionaries relating physics concepts and interactions to some  necessary conditions for life, but they cannot encompass sufficient conditions for life to exist. They are maps from some aspects of physics to some aspects of biology but they omit essential aspects of what it is to be alive, 
 failing to capture the nature of biology as a whole, because they miss essential biological features. Equivalently, 
 there are essential biological variables that cannot be captured by such a dictionary.} 
		}
	\end{quote}
From a physics viewpoint, this claim is justified by discussion in \cite{Anderson and Stein 1987} regarding the contrast between equilibrium systems, dissipative systems, and life.

\subsection{Strong Emergence in Biology}\label{sec:Strong_Emerge_Biol}

There are many reasons why biology is strongly emergent. I note first that biological emergence is based in \href{https://en.wikipedia.org/wiki/Supramolecular_chemistry}{supramolecular chemistry} \cite{Lehn 1993 } \cite{Lehn 1995 } and the nature of biomolecules such as \href{https://en.wikipedia.org/wiki/Nucleic_acid}{nucleic acids} and \href{https://en.wikipedia.org/wiki/Protein}{proteins} \cite{Petsko_Ringe_2009}. Given this understanding, reasons biology is strongly emergent are as follows:
\begin{itemize}
	\item 
	Firstly, because organisms are made of biomolecules where the functional  properties of those molecules  \textbf{P} are strongly emergent due to their dynamic symmetry-breaking nature, see Section \ref{sec:biology_diachronic}. 
	
	\item Secondly, because of the functional emergence \textbf{P}  of the properties of physiological process and  organisms at any specific time out of the underlying biomolecules. These are the processes of \href{https://en.wikipedia.org/wiki/Physiology}{physiology} \cite{Physiology_human}, which are dynamic processes based in essentially biological interactions and variables. 
	
	\item Thirdly, because of the  emergence \textbf{E}  of each individual  organism over time out of the component biomolecules through the processes of \href{https://en.wikipedia.org/wiki/Developmental_biology}{developmental biology} \cite{Carroll 2005} \cite{Carroll 2008} 
	controlled by irreducible hierarchical gene regulatory networks.
	
	\item Fourthly, because of the historical emergence \textbf{E} over evolutionary timescales of organisms, of the biomolecules out of which they are made \cite{Wagner 2014}, including the DNA that is our genetic material \cite{Mayr 2001}, and of the developmental processes whereby biomolecules give rise to organisms. This is an intricate intertwined process of evolution and development (hence \href{https://en.wikipedia.org/wiki/Evolutionary_developmental_biology}{ \textit{Evo-Devo}} 
	\cite{Carroll 2008},
	 \cite{Oyama Griffiths Cycles}). Biology has an ineliminable historical element which results in its present day nature through environmental adaptation, a top-down process \cite{Campbell_1974}.	
\end{itemize}
\noindent All these dimensions of biology are strongly emergent.  
\begin{quote}
	\textbf{Conclusion}: \textit{\textbf{Biology is not reducible to physics. It is an indubitable case of strong emergence of properties \textbf{P} because it  involves function and purpose (\cite{Hartwell_et_al_1999} associated with downward causation and information flows \cite{Nurse 2008} \cite{Davies 2019} implemented by time dependent constraints \cite{Ellis_Kopel_2019}, and because it involves  strong emergence associated with molecular biology and with physiology. Its emergence \textbf{E} is also strong  emergence because of the nature of the process of Natural Selection that led to life on Earth \cite{Mayr 2001}, which involves both randomness and downward causation  \cite{Campbell_1974}.}}
\end{quote}
This emergence results in testable Effective Theories, such as  the Hodgkin-Huxley equations for action potential propagation, Fitness Landscapes \cite{Wright 1932} \cite{McGhee 2006}, models of the heart \cite{Fink_and_Noble}, and Darwinian evolution \cite{Mayr 2001}. 
\section{Conclusion}\label{sec:conclude}


The discussion above has highlighted that while one can indeed obtain Effective Theories of almost any topic studied by physics --- indeed that is what physics is about --- one can almost always not make the link between the micro and emergent theories without explicitly introducing variables and concepts that were not implied by the underlying theory. A purely bottom-up attempt of derivation of the properties of emergent states such as semi-conductors cannot be done: these properties are strongly emergent.
Downward causation \cite{Ellis_2016} plays a key role in strong emergence in condensed matter physics and soft matter physics (Section \ref{sec:Strong_emerge}), even if it is not explicitly identified as such. 
 
Biology is strongly emergent for multiple reasons (Section \ref{sec:Strong_Emerge_Biol}). The Darwinian process of natural selection that leads to the existence of particular living forms, is strongly environmentally dependent, and hence unpredictable in a bottom up way. The  processes of molecular biology are shaped in a downward way by signalling molecules that change protein conformation and hence alter constraints on ion and electron motion in a time dependent way, related to biological purpose and function.

I now comment on whether micro data dependence undermines strong emergence (Section \ref{sec:determine}), the nature of essentially higher level variables (Section \ref{sec:higher_var}), whether effective field theories disprove strong emergence (Section \ref{sec:EFTs}), 
and the equal validity of all emergent levels (Section \ref{sec:equal_valid}). Section \ref{sec:novel}  summarise results obtained in this paper.

\subsection{Does micro data dependence undermine strong emergence?}\label{sec:determine}

It has been strongly claimed to me that micro data dependence of all outcomes undermines my argument for strong emergence.  The argument goes as follows. 
Suppose I am given the initial positions $\textbf{r}_i$ and momenta $\textbf{p}_i$ of all particles in the set 
${\cal P}$
\begin{equation}\label{eq:L1}
{\cal P} := \textrm{(protons, neutron, electrons)}
\end{equation}
   at a foundational level \textbf{L1}, which constitute an emergent structure \textbf{S} at a higher level \textbf{L2}. The details of \textbf{S} are determined by that microdata, even though its nature cannot be recognised or described at level \textbf{L1}. The forces between the particles at level \textbf{L1} completely determine the  dynamics at that level. Hence the emergent outcomes at level \textbf{L2} are fully determined by the data at Level \textbf{L1}, so the emergence of dynamical properties and outcomes at level \textbf{L2} must be weak emergence and be predictable, at least in principle from the state (\ref{eq:L1}) of level \textbf{L1}.  This would apply equally to physical, engineering, and biological emergent systems. It is in effect a restatement of the argument from supervenience.\\

\noindent There are problems with the argument just stated as regards microphysics, macrophysics, biology, and causal effectiveness of higher levels.

\paragraph{Microphysics} 
 There is irreducible uncertainty of quantum outcomes when wave function collapse to an eigenstate takes place, with outcomes only predicted statistically via the Born rule \cite{Ghirardi}. This unpredictability has 
consequences that can get
amplified to macrolevels, for example in terms of causing errors in computer memories due to cosmic rays. Cosmic rays also alter genes significantly enough to cause cancer and to alter evolutionary history (\cite{Ellis_closure} \S 7.2). Quantum uncertainty also  affects specific structure formation outcomes in inflationary cosmology (\cite{Ellis_2016}:p.398).    

\paragraph{Macrophysics}
Consider a higher physical level \textbf{L3} in the context of a fluid where 
 convection patterns take place. Because of the associated chaotic dynamics together with the impossibility of setting initial data to infinite precision \cite{Ellis Meissner Nicolai}, macroscopic outcomes are  unpredictable in principle from micro data (\ref{eq:L1}) \cite{Gisin_2019}.   \cite{Anderson_01} puts it this way: 
\begin{quote}
	``\textit{A fluid dynamicist when studying the chaotic outcome of convection in a Benard cell knows to a gnat's eyelash the equations of motion of this fluid but also knows, through the operation of those equations of motion, that the details of the outcome are fundamentally unpredictable, although he hopes to get to understand the gross behaviour. This aspect is an example of a very general reality: the existence of universal law does not, in general, produce deterministic, cause-and-effect behaviour}''
\end{quote}
This fundamentally undermines the concept of a causally closed  physical levels in the case of classical physics. This affects local fluid convection patterns as well as weather patterns, from where it chains down to affect macro biology, microbiology, and thus the underlying physics. Because of global warming, this is causing serious problems for farming.\footnote{See \href{https://www.dw.com/en/climate-change-and-farming-unpredictability-is-here-to-stay/a-44836642}{Climate change and farming: 'Unpredictability is here to stay'}.} 

Furthermore chaotic motion is unpredictable because of gravitational effects.   Michael Berry has pointed out that after about 120 collisions a billiard ball's future behavior would be effected by the gravitational pull of one electron at the edge of the universe.\footnote{I thank Martin Rees for that comment.} 

\paragraph{Biology: 
 Harnessing stochasticity} Consider the case of bio molecules in a cell where a molecular storm occurs \cite{Hoffman}. In talking about the molecular machines that power cells, he states that every such molecular machine in our bodies is hit by a fast-moving water molecule about every $10^{-13}$ seconds. ``\textit{At the nanoscale, not only is the molecular storm an overwhelming force, but it is also completely random.}'' To extract order from this chaos,  ``\textit{one must make a machine that could `harvest' favorable pushes from the random hits it receives.}'' That is how biology works at this level. By selection for biological purposes from an ensemble of possibilities presented by this randomness, higher level needs may be selected for in a downward way, as stated by \cite{Noble and Noble 2018 Stochasticity}:
\begin{quote}
	``\textit{Organisms can harness stochasticity through which they can generate many possible solutions to environmental challenges. They must then employ a comparator to find the solution that fits the challenge. What therefore is unpredictable in prospect can become comprehensible in retrospect. Harnessing stochastic and/or chaotic processes is essential to the ability of organisms to have agency and to make choices.} 
\end{quote}
This is the opposite of the Laplacian dream of ordered initial data at the micro level leading to outcomes uniquely predicted by that micro data. It is the detailed structure of molecular machines, together with the lock and key molecular recognition mechanism used in molecular signalling \cite{Berridge}, that enables the logic of biological processes to emerge as effective theories governing dynamics at the molecular level and thereby reaching down to control the physical level \textbf{L1} via time dependent constraints \cite{Ellis_Kopel_2019}.

\noindent In this way downward causation by irreducible higher level variables takes place, and alters lower level dynamics so that the initial data does not determine outcomes, it is higher level needs that do so.  
But the issue then is, 
 what is the nature of such variables? Do they indeed exist? I look at this in Section \ref{sec:higher_var}.

\paragraph{Causal effectiveness of higher levels} Finally, there is what is in effect another version of this argument, this time in the backwards direction of time. The challenge is to explain how  life as it  exists nowadays arose out of the initial data in the early universe  (to be simple, let's say data on the last scattering surface where matter and radiation decoupled from each other). It could not have arisen in a deterministic way from that initial data because of all the uncertainties and randomness just discussed. The obvious explanation is that it arose through the combination of evolutionary and developmental process (`Evo-Devo' \cite{Carroll 2005} \cite{Carroll 2008}) which are essentially biological processes which have real causal power at their emergent level, enabled but not determined by the underlying physics (see Section \ref{sec:natural_select_emerge}). 

Martin Rees  however says in relation to embryo development\footnote{Private communication}, \begin{quote}
 	``\textit{It could be the case that if we `run the film backwards'
 through successive hyper-surfaces back to the big bang, every event or
state is emergent from its predecessors via a combination of micro
events (like a breaking wave) even though we interpret it in terms of
all the higher levels. ... you could trace back across `hypersurfaces'  to learn how viewed on an atomic
scale, the embryo had evolved into the adult''}
 \end{quote} 
As phrased, the underlying view is that both the evolutionary and developmental processes are epiphenomena floating on the surface of the real causal element, namely the atomic level physics operating on data (\ref{eq:L1}). The underlying viewpoint is that all causation proceeds upwards in the hierarchy of complexity. 

Of course you could trace the micro events in this way. But this would simply fail to tell you what was driving these developments: what was the key causal factor shaping these outcomes? You can do that micro  level tracing whether the higher levels have zero causal powers, as is suggested by this phrasing, or they in fact have real causal powers because every level has causal powers (\cite{Noble 2012} and  Section \ref{sec:equal_valid}). 
 One either has to claim that genes and gene regulatory networks have real causal powers \cite{Watson et al 13} \cite{Alberts_etal_07}, as has apparently been demonstrated by many thousands of experiments, or that all these experimentally determined outcomes are just correlations not indicating causation because no causal powers reside at either the molecular or cellular level. 
 
 Tracing back what interactions took place at the atomic level as suggested by Rees has no ability to distinguish between these two fundamentally different possibilities. It simply records what happened but not why it happened. 
My position, following \cite{Campbell_1974} and \cite{Noble2008_Music} is that the higher levels have real causal powers based  on essentially biological processes \cite{Hartwell_et_al_1999} \cite{Campbell and Reece 2008} \cite{Nurse 2008} enabled by downward causation, for which their is ample experimental and counterfactual evidence (vary higher level variables and observe or argue for altered lower level outcomes). 

\subsection{Essentially higher level variables}\label{sec:higher_var}

 Irreducible higher level variables cannot be obtained by coarse graining or in any other way upscaling lower level variables, that is, they are essentially higher level variables defined at a level $\textbf{L} > \textbf{L1}$.  There are many examples 
\begin{itemize}
	\item \textbf{Structures }\cite{Green and Batterman 2017} \cite{Batterman 2018} give examples involving heterogeneous physical structures and biology.   \cite{Davidson et al 2009} give an example from developmental biology, the notochord: 
	\begin{quote}
		``\textit{The capacity of the notochord to resist bending as it extends the embryo comes from the
			structure of the whole notochord. Measurements at the level of the individual collagen fiber
			or fluid-filled cells that make up the structure would not reveal the mechanical properties of
			the whole notochord}.''
	\end{quote}
This factor plays a key role in developmental dynamics, thereby affecting all levels down to \textbf{L1}.  \cite{Anderson and Stein 1987} cite rigidity as an example of true emergence: \textit{``We emphasize that this rigidity is a true emergent property: none of the forces between actual particles are capable of action at a distance. It implies that the two ends cannot be decoupled completely without destroying the molecular order over a whole region between them.''}  

%
\item \textbf{Feedback system goals}  
An important case is feedback control \cite{Wiener 1948} (engineering), which is essentially the same as homeostasis \cite{Guyton 1977} (biology) , implemented by a feedback control loop (\textbf{Figure 2}). This is one of the network motifs characterised by \cite{Alon 2006}. 
Then the constraints  ${\cal C}_\textbf{LI}(t)$ in (\ref{eq:Laws_of_biology}) depend on goals $G_ \textbf{LI}$ valid at level \textbf{L} but set at the Level of Influence \textbf{LI}. The goals $G_ \textbf{LI}$ are irreducible higher level variables because they obtain their causal power only by their undeniable effects via this closed loop. 
\begin{figure}[h]
	\centering
	\includegraphics[width=0.45\linewidth]{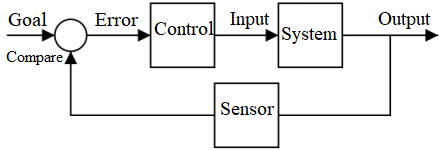}
	\caption{\textbf{Feedback control loop} \textit{A goal determines outputs by being compared with the current system state, the difference being fed back to a controller that alters the state of the system (this is an essentially non-linear relation, which is why it cannot be characterised by any simple aggregation process).}}
	\label{fig:simplefeedbackcontrolloop2}
\end{figure}

An engineering example is adjusting the (macro level) temperature setting on a thermostat, which determines the (micro level) motions of billions of molecules. The goals $G_ \textbf{LI}$ are irreducible macro level variables (their causal power vanishes if the feedback loop is broken). 
 It is the dynamics at the feedback control loop level \textbf{LI} that determines what happens at that level, and hence at lower levels; this is the element of downward causation, where cause is understood as in \cite{Pearl 2009} \cite{Pearl_Why}. This is a special case of a network.

\item \textbf{Networks} Green and Batterman support the argument here by the general example of network dynamics. They say  \cite{Green and Batterman 2020} 
\begin{quote}
	``\textit{The characteristics of functional equivalence
		classes are explained with reference to how network structures constrain dynamic outputs to enable
		generic types of functions such as sustained oscillations, noise filtering, robust perfect adaptation,
		signal amplification, etc.}''
\end{quote}  
These are functional principles that affect what networks actually exist. They lead to network motifs \cite{Alon 2006} that represent higher level organising principles independent of the lower level dynamical details, that occur in many networks. 

\item \textbf{Dynamical system attractors} %
Consider a physical level \textbf{L3} whose dynamics can be characterised as a dynamical system \cite{Arnold}.
It will have attractors at that level 
that are the causal element that order dynamic outcomes. They cannot in principle be characterised at Level \textbf{L1}, for they essentially involve the macro level interactions,  but  they reach down to shape dynamics at level \textbf{L1}, as for example in the case of fluid convection \cite{Bishop_2008_convection}. 
Paradoxically, they are not emergent rules, i.e. rules arising out of the lower level dynamics. Rather they are rules that characterise higher level dynamics of emergent entities  
quite independently of lower level dynamics \cite{Green and Batterman 2020}.
 This is parallel to the claim by  \cite{Laughlin_Pines_2000} of existence of classical and quantum protectorates, governed by dynamical rules that characterise emergent systems as such.

\item \textbf{Biological Needs} 
 affect relative reproductive success, with selection reaching down to shape lower level outcomes \cite{Campbell_1974} so as to meet  higher level needs according to Eqn.(\ref{eq:adaptive_select}). The selection criteria $c_\textbf{LI}$ are higher level criteria related to biological purpose and function (Section \ref{sec:natural_select_emerge}). In effect, needs such as the ability to move  and see act as attractors - higher level organising principles - for biological macro-evolution. This then reaches down to shape lower levels \cite{Campbell_1974}, with these needs leading to convergence to one or other of the restricted set of structures that can realise the needed function \cite{McGhee 2006}.  


\end{itemize}
Underlying all these examples is the fact that Effective Theories $\textbf{EF}_\textbf{L}$ at each level \textbf{L} are indeed effective.  
This is discussed further in \S \ref{sec:equal_valid}.

\subsection{Do Effective Field Theories disprove Strong emergence?}\label{sec:EFTs}
It has been suggested by \cite{Hossenfelder_2019} and \cite{LM} that Effective Field Theories 
(EFTs) rule out strong emergence because they  enable derivation of  emergent properties \textbf{P} in a bottom up way,\footnote{Note that `bottom-up' and 'top-down' are used
	exactly in the opposite way in discussions about EFTs and emergence. The top-down construction
	of an EFT means that we are constructing the EFT by deriving it from a more fundamental high energy
	theory. I will use the emergence appellation.}  in Hossenfelder's case even saying one should be able to in principle deduce election results in this way.

There is a wide variety of EFTs \cite{Hartmann 2001}, with some being closely related to the Renormalisation Group \cite{Castellani_02} \cite{Burgess_2007}. The basic idea is to use a power series expansion of  the Lagrangian in terms of $1/M$ where $M$ is the relevant mass scale of the microscopic theory, using symmetry principles to constrain the expansion, and using a power counting scheme. Using the EFT allows one to ignore
the dynamics of fields relevant at high energies.
  
 The philosophical literature on the topic includes \cite{Bain 2013}, \cite{Butterfield 2014}, and \cite{Rivat and Gribaum 2020}, who  comment on the use of EFTs in condensed matter physics:
\begin{quote}
	``\textit{Yet another example is the use of EFTs in
	condensed matter physics: even when the underlying theory is known,
	often the only tractable way to compute low-energy observables is to
	build an effective model as if the underlying theory were unknown}''.
\end{quote}
In other words it is derived as an EFT at the emergent level, without trying to derive it from the more fundamental level. Hence  in this case it has nothing to say about emergence. 

Most importantly, \cite{Bain 2013} claims that if two successive theories related by an EFT are structurally different, for example if the Lagrangian has different symmetries, then the lower energy theory is essentially different  from the underlying theory. He proposes
\begin{description}
\item[(a)] \textbf{Ontological dependence}. \textit{Physical systems described by an EFT are ontologically
	dependent on physical systems described by a high-energy theory.}
	\item[(b)] \textbf{Failure of law-like deducibility}. \textit{If we understand the laws of a theory encoded in a
Lagrangian density to be its Euler-Lagrange equations of motion, then the phenomena
described by an EFT are not deducible consequences of the laws of a high-energy theory.}
\item[(c)] \textbf{Ontological distinctness}. \textit{The degrees of freedom of an EFT characterize physical systems
that are ontologically distinct from physical systems characterized by the degrees of
freedom of a high-energy theory.}
\end{description}
Thus Bain's paper supports both epistemological and ontological strong emergence. This viewpoint  is controversial \cite{Rivat and Gribaum 2020}, but it suffices to prove that one cannot in an easy way state that EFTs disprove strong emergence. His view is supported by the fact that the broken symmetries result in the emergent state having an orthogonal Hilbert space to the Hilbert space of the underlying theory \cite{Anderson 1984}. That is why they can be claimed to be ontologically different. 

 In any case \cite{Adams_et_al_2006} show that some apparently perfectly
sensible low-energy effective field theories governed by local, Lorentz-invariant Lagrangians, are
secretly non-local, do not admit any Lorentz-invariant notion of causality, and are incompatible
with a microscopic S-matrix satisfying the usual analyticity conditions.
There are pitfalls in the way of claiming EFTs can characterise emergence of arbitrary low energy phenomena. 

\paragraph{Renormalisation group approaches} EFTs are often closely related to Renormalisation Group (RG) approaches.  \cite{Butterfield 2014} argues that the renormalisation group is compatible with Nagelian reduction, where as he explains, that idea has been extended to entail as regards the relation of the reducing theory T1 to the emergent theory T2  that
\begin{quote}
	(i) T2 may well have predicates, or other vocabulary, that do not occur in T1. So
	to secure its being a sub-theory, we need to augment T1 with sentences introducing
	such vocabulary, such that T2 is deducible from T1 as augmented.
\end{quote}
Thus in my terms (Section \ref{sec:weak_strong}), this is a case of weak derivation and hence a case of strong emergence. This is reinforced by the further extension \cite{Butterfield 2014}
\begin{quote}
	(ii) What is deducible from T1 may not be exactly T2, but instead some part,	or close analogue, of it. There need only be a strong analogy between T2 and what strictly follows from T1. Nagel called this approximative reduction.
\end{quote}
This is simply not a genuine upward derivation of the emergent phenomena. It agrees with what \cite{Leggett_1992} claims (see \S  \ref{sec:bottom_up_problems}) and is compatible with strong emergence as defined by  \cite{Chalmers_2000}.  Furthermore 
\cite{Green and Batterman 2020} claim, on the basis of the universality classes that emerge,  that a renormalisation group explanation  extracts structural features that
stabilize macroscopic phenomena irrespective of changes in or perturbations of microscopic
details, hence supporting strong emergence. 
\cite{Morrison 2012} strongly makes the same point:   
\begin{quote}
	``\textit{What is truly significant about 
		emergent phenomena is that we cannot appeal to microstructures in explaining or predicting these phenomena, even though they are constituted
		by them. Although this seems counter intuitive, RG methods reveal the nature of this ontological independence by demonstrating (1) how systems
		with completely different microstructures exhibit the same behavior, (2) how successive transformations give you a Hamiltonian for an ensemble that contains very different couplings from those that governed the initial ensemble, and (3) the importance of the physics behind the
		notion of an infinite spatial extension for establishing long-range order. ... 
RG reveals why the microdetails of symmetry breaking are
		neither ontologically nor epistemically necessary for emergence; the information is simply lost as the length scale changes.}''
\end{quote}   
This argument goes even further than the symmetry  breaking arguments  given above.
 
\subsection{Equal Validity of Levels}\label{sec:equal_valid}

Luu and Meissner in a paper on Effective Field Theories \cite{LM} state that in physics, one has an ``\textit{Equal validity of all levels}''. This is indeed the case: this is the physics version of Denis Noble's ``Principle of Biological Relativity'' \cite{Noble 2012},  which states that in biology, no emergent level is privileged over any other.

This is expressed very nicely by Sylvan Schweber in \cite{Schweber 1993}, commenting on Phil Anderson's views:
\begin{quote}
	\textit{``Anderson  believes in  emergent laws.   He holds the
		view that each level has its  own ``fundamental'' laws and its own ontology. Translated into the language of particle physicists, Anderson would say each level has its effective Lagrangian and its set of quasistable particles. In  each level the effective Lagrangian - the ``fundamental'' description at that level -  is  the best  we can  do.'' 
	}
\end{quote}
This has to be the case because we don't know the underlying Theory of Everything (TOE) of physics, if there is one, and so don't - and can't - use it in real applications. So all the physics laws we use in applications are effective laws in this sense, applicable at the appropriate level (Section \ref{sec:P(d)_solid_state}). It may be the Standard Model of Particle Physics, or Quantum Field Theory, or Quantum Theory, or nuclear physics, or atomic physics, and so on, depending on the problem at hand; each is a very well tested effective theory at the appropriate level. Similarly, there are very well tested effective theories at each level in biology: the molecular level, the cellular level, the physiological systems level for example. 

The point then is this: each of them represents a causally valid theory, where causation can be defined as in \cite{Pearl 2009} \cite{Pearl_Why}, that holds at its level.  None of them can be deemed to be more or less fundamental than any other, because \textit{none of them is fundamental} (i.e. none is the hoped for TOE).
   
In terms of Effective Theories \cite{Castellani_02},  this is the statement  
\begin{quote}
	\textbf{Equal Causal Validity}: \textit{\textbf{Each well established  emergent level \textbf{L} in physics and biology represents an Effective Theory   $\textbf{ET}_\textbf{L}$  as represented in (\ref{eq:effective_laws}), with variables $v_\textbf{L}$ appropriate to that level, so each level is equally valid with each other in a causal sense. None is the fundamental theory to which all else reduces}.}
\end{quote}
In particular, there is no known bottom-most physical level to which all of physics - or any other emergent level - can be reduced. No TOE can invalidate causation taking place at each emergent level. 
This equal causal validity occurs because  higher levels are linked to lower levels by a combination of upwards and downwards  causation (sees Section \ref{sec:Strong_emerge} and \cite{Noble 2012}, \cite{Ellis_2016}). 
 I develop this all in more detail in a companion paper on interlevel  causal closure \cite{Ellis_closure}. The punchline is that  whatever we may consider as the bottom-most physics level, it is not causally closed by itself. One must take into account higher levels in order for causal closure to occur.

\subsection{Novel Results}\label{sec:novel}  

New results in this paper include, 
\begin{itemize}
	\item The distinguishing in Section \ref{sec:kinds_of_symmetry_breaking} of three different kinds of symmetry breaking mechanisms: \textbf{SSB}(\textbf{m}), Spontaneous Symmetry Breaking occurring at the micro level; \textbf{SSB}(\textbf{M}), Spontaneous Symmetry breaking occurring due to the emergence processes $\textbf{E}$ creating  the macro level from the micro  level, and  \textbf{SB}(\textbf{NS}), symmetry breaking  due to Darwinian processes of natural selection.
	\end{itemize}
	On this basis, I provide 
	\begin{itemize}
	\item The argument in Section \ref{sec:SSB(M)} that emergence of properties based in \textbf{SSB}(\textbf{M}) is strong emergence. Such broken symmetries are the core of condensed matter physics \cite{Anderson_1981_symmetry}, \cite{Anderson_89}. 
	 This implies solid state physics properties \textbf{P} are strongly emergent.
		\item The argument in Section  \ref{sec:downward_diachronic_synchronic} that lower level physics \textbf{m} is causally incomplete because it cannot by itself produce experimentally established outcomes of solid state physics. It only becomes causally complete if variables \textbf{a} based in higher level conditions are included to produce  effective lower level dynamics \textbf{m'} that give the correct higher level outcomes, i.e. if downward causation $\textbf{m} \rightarrow \textbf{m'(\textbf{m,a})}$ takes place. I develop the issue of causal closure in depth in a companion paper \cite{Ellis_closure}.  
	\item The argument in Section \ref{sec:Strong_Emerge_Biol} that strong emergence takes place in the case of living systems due to both \textbf{SSB}(\textbf{M}) and \textbf{SB}(\textbf{NS}).
	\item The argument in Section \ref{sec:equal_valid}) that all levels are equally causally effective. 
	\item Supporting arguments in Section \ref{Bio_variables} showing there is no dictionary relating basic physics variables to biology, and   in Section \ref{sec:higher_var} showing that essentially higher level variables do indeed occur. 
	\item Section \ref{sec:determine} shows  that detailed initial data at the micro level does not uniquely determine outcomes at later times, allowing space for higher level variables to do so,  and Section \ref{sec:EFTs} shows that  Effective Field Theories do not disprove strong emergence.   
\end{itemize}

\paragraph{Acknowledgments}  I thank Barbara Drossel, Andrew Briggs, and Martin Rees  for  helpful comments. I thank Carole Bloch and Rob Adam for   extremely helpful proposals regarding the scope of this paper. I thank two anonymous referees for comments that have greatly strengthened the paper.

\pagebreak


\noindent Version 2020/06/30


\end{document}